\documentclass[preprint]{autart} 

\usepackage{graphicx}          
\usepackage{subcaption}
\usepackage{derivative}
\usepackage{amsmath,mathtools}
\usepackage{xfrac}
\usepackage{physics}
\usepackage[numbers,sort&compress]{natbib}
\usepackage{enumitem} 
\usepackage{xcolor}
\usepackage{amsfonts}
\usepackage{algorithm}
\usepackage{algpseudocode}
\usepackage{float}

\begin{document}

\begin{frontmatter}

\title{Geometric Reachability for Attitude Control Systems via Contraction Theory}


\author[ISEE]{Chencheng Xu}\ead{xucc@zju.edu.cn},
\author[Saber]{Saber Jafarpour}\ead{saber.jafarpour@colorado.edu},
\author[CSE]{Chengcheng Zhao}\ead{chengchengzhao@zju.edu.cn}, 
\author[ISEE]{Zhiguo Shi}\ead{shizg@zju.edu.cn},
\author[CSE]{Jiming Chen}\ead{cjm@zju.edu.cn}

\address[ISEE]{College of Information Science and Electronic Engineering, Zhejiang University, Hangzhou, China.}  
\address[Saber]{Department of Electrical, Computer, and Energy Engineering, University of Colorado Boulder, Boulder, CO, USA.}  
\address[CSE]{College of Control Science and Engineering, Zhejiang University, Hangzhou, China.}  


\begin{abstract}
    In this paper, we present a geometric framework for the reachability analysis of attitude control systems.  
We model the attitude dynamics on the product manifold $\mathrm{SO}(3) \times \mathbb{R}^3$ and introduce a novel parametrized family of Riemannian metrics on this space.  
Using contraction theory on manifolds, we establish reliable upper bounds on the Riemannian distance between nearby trajectories of the attitude control systems.  
By combining these trajectory bounds with numerical simulations, we provide a simulation-based algorithm to over-approximate the reachable sets of attitude systems.  
We show that the search for optimal metrics for distance bounds can be efficiently performed using semidefinite programming.  
Additionally, we introduce a practical and effective representation of these over-approximations on manifolds, enabling their integration with existing Euclidean tools and software.  
Numerical experiments validate the effectiveness of the proposed approach.
\end{abstract}

\end{frontmatter}

\section{Introduction}
Recent advances in technology have significantly enhanced the capabilities and performance of autonomous systems, such as drones, spacecraft, and robotic arms.
However, despite these significant benefits, safety and reliability considerations remain one of the main challenges in their control and coordination.
Many established theoretical works have successfully addressed safety-critical tasks for system states evolving in Euclidean space.
For example, classical obstacle avoidance scenarios focus on the translational states of these systems. 
Reachability analysis has been employed in this context to address safety problems in the space of translational motion~\cite{djeumou2022fly,she2020over,lew2021sampling}, which generates a certifiable region that contains all possible states a system can reach from a specified initial set.
On the other hand, specialized safety-critical tasks such as aggressive maneuvering~\cite{liu2018search}, active vision~\cite{falanga2017aggressive,abate2023run}, optical shielding, and communication~\cite{tan2020constrained,danielson2021spacecraft} underscore the need to study the reachability of rotational motion and attitude dynamics of these autonomous systems. 

The configuration space of a rigid body's rotational motion is the set of $3 \times 3$ orthogonal matrices with a determinant of 1, known as the special orthogonal group $\mathrm{SO}(3)$.
Although this space is non-Euclidean, numerous studies in the literature propose alternative parametrization to model attitude dynamics in an Euclidean formulation~\cite{shuster1993survey,chaturvedi2011rigid}. 
This leads to the parametrization-based reachability concept, offering a practical approach to representing and analyzing attitude dynamics within Euclidean charts, and thereby enabling the application of Euclidean reachability techniques. 
These reachability approaches can be categorized into Hamilton–Jacobi formulations~\cite{zagaris2018applied,sun2019quadrotor,chen2018hamilton}, set-propagation methods~\cite{lang2021formal,huang2022polar,althoff2021set}, and simulation-based methods~\cite{huang2012computing,duggirala2013verification,fan2017simulation}.
Among these, simulation-based methods have proven effective for systems with complex, high-dimensional nonlinear dynamics~\cite{girard2006verification,donze2007systematic,maidens2014reachability,fan2017simulation}.
These methods over-approximate the reachable set by using a finite number of numerical simulations, making them computationally efficient and scalable to high-dimensional state spaces.
Euclidean contraction theory~\cite{lohmiller1998contraction,sontag2010contractive,tsukamoto2021contraction} is effectively applied in this context to measure the convergence of nearby trajectories and establish reliable upper bounds on the distance between them.
However, none of these parametrization-based methods provide a global analysis across the entire attitude space. In fact, challenges such as singularities and the unwinding phenomenon~\cite{chaturvedi2011rigid} can arise in these representations, limiting the effectiveness of Euclidean reachability approaches.
These observations underscore the need to develop a global, coordinate-free approach for reachability analysis in attitude control systems.

Geometric methods have shown significant promise in the verification and synthesis of systems evolving on non-Euclidean spaces, as they utilize a coordinate-free formulation of the problem and avoid singularities. They have been successfully applied to study certain notions of safety in attitude control systems. 
For example, geometric barrier certificates~\cite{wu2015safety,xu2023formal} are introduced to encode attitude specifications, which are defined on $\mathrm{SO}(3)$, into control barrier functions. This allows for the design of optimal and safe attitude controllers~\cite{wu2015safety,tan2020construction, tan2021high} and safety verification for attitude control systems~\cite{xu2023formal}.
Another relevant literature is the work~\cite{shafa2024guaranteed} that under-approximates the reachable set of unknown systems on manifolds using their Lipschitz bounds. 

In this paper, we adopt a geometric approach to efficiently compute over-approximations of reachable sets for attitude control systems. 
We model the attitude dynamics on the product manifold $\mathrm{SO}(3) \times \mathbb{R}^{3}$, where we establish a simulation-based framework to over-approximate the reachable sets by combining provable upper bounds on trajectory distances and numerical simulations. 
%
We introduce a novel family of left-invariant Riemannian metrics, parameterized by positive definite matrices, on the product manifold $\mathrm{SO}(3) \times \mathbb{R}^{3}$. This metric structure provides suitable properties for characterizing distances and representing sets on this manifold.
Using the theory of contracting systems on manifolds, we establish provable upper bounds for the incremental distance between attitude trajectories, which can be effectively characterized by linear matrix inequalities.
Metric balls of this left-invariant Riemannian structure play a crucial role in over-approximating reachable sets of the attitude dynamics. Although providing an explicit expression for the distance functions and metric balls in these spaces is complicated, we provide easy-to-check conditions for verifying set inclusions in the metric balls.
Building on these geometric results, we develop a novel simulation-based algorithm for reachability analysis on the manifold $\mathrm{SO}(3) \times \mathbb{R}^{3}$. We apply a geometric decomposition and partitioning approach on $\mathrm{SO}(3)$ to generate parallelizable simulations for reachability analysis. 
We demonstrate that determining the optimal metrics for upper bounds on distances between reachable and simulation states can be efficiently formulated and solved using semidefinite programming (SDP) problems. This approach can then be effectively applied throughout our simulation-based algorithm to generate a locally tight approximation.
Furthermore, we introduce an equivalent and practical representation of metric balls within a special Euclidean atlas on $\mathrm{SO}(3)$, which enables further analysis and visualization of the reachability results using classical Euclidean tools and software.
 
The rest of the paper is organized as follows. Section~\ref{Sec:Pre_Prob} introduces fundamental concepts in differential geometry and presents the problem formulation. In Section~\ref{Sec:Contrac_thm}, we introduce a class of left-invariant Riemannian metrics on $\mathrm{SO}(3) \times \mathbb{R}^{3}$ and demonstrate how contraction theory is extended and applied to this manifold. Section~\ref{Sec:Reach_methods} presents the geometric framework for simulation-based reachability and Section~\ref{Sec:Metric_Ball_Representation} discusses how the results can be effectively represented in a Euclidean atlas. Finally, Section~\ref{Sec:Num_examples} provides a numerical example to validate the proposed method.

\section{Preliminaries and Problem Formulation}\label{Sec:Pre_Prob}
The set of real numbers and the set of non-negative real numbers are denoted by $\mathbb{R}$ and $\mathbb{R}_{\ge 0}$, respectively. Matrices are represented by uppercase letters (\textit{e.g.} $Q$), vectors in $\mathbb{R}^{3}$ are denoted by bold lowercase letters (\textit{e.g.} $\boldsymbol{x}$), and the entries of vectors and matrices are indicated using non-bold letters with subscript indices (\textit{e.g.} $x_{i}$ and $Q_{ij}$). The standard $\ell_2$-norm on $\mathbb{R}^{3}$ is denoted by $\left \| \cdot \right \|$. We use $I$ to denote the $3 \times 3$ identity matrix. 

\subsection{Differential geometry}
Let $M$ be a $m$-dimensional smooth manifold and let $T_{p}M$ denote the tangent space at $p$. The tangent bundle is $TM = \left \{ (p,v) \mid p \in M, v \in T_{p}M\right \}$. A vector field $V$ on $M$ is a map that assigns to each point $p \in M$ a tangent vector $V(p) \in T_{p}M$. The set of all vector fields of class $C^{\infty}$ on $M$ is denoted by $\mathcal{X}(M)$.  
Given a smooth function $f: M \to \mathbb{R}$, the Lie derivative of $f$ with respect to the vector field $V$, denoted by $V(f):M \to \mathbb{R}$, is defined as $V(f)(p) = V(p)(f) = \lim_{t \to 0} \frac{f(p + tV(p)) - f(p)}{t}$, for each $ p \in M$.
A Riemannian metric $\mathbb{G}$ on the manifold $M$ defines an inner product at each point $p \in M$, denoted by $\langle \cdot,\cdot \rangle_{\mathbb{G}(p)}: T_{p}M \times T_{p}M \to \mathbb{R}$.
This Riemannian structure introduces a notion of length for curves and a notion of distance between points. Consider a curve $\gamma:[a,b] \to M$ on the Riemannian manifold $( M,\mathbb{G})$. The length of this curve is given by $ \mathcal{L}_\gamma = \int_{b}^{a} \sqrt{\langle \dot{\gamma}(t),\dot{\gamma}(t) \rangle_{\mathbb{G}(\gamma(t))}}{\rm d}t$. A distance function $d:M \times M \to \mathbb{R}_{\ge0}$ gives the length of the shortest path between two points, which is equivalent to the length of the minimal geodesic connecting them. 
Let $\nabla: \mathcal{X}(M) \times \mathcal{X}(M) \to \mathcal{X}(M)$ represent an affine connection on $M$. For any smooth function $f: M \to \mathbb{R}$ and vector fields $V, W \in \mathcal{X}(M)$, the following holds:
\begin{align}
    \nabla _{(fV)}W &= f \nabla_{V}W, \label{Connection_prop_1} \\
    \nabla_V (fW) &= V(f)W + f\nabla _{V}W, \label{Connection_prop_2}
\end{align}
A connection $\nabla$ is called the Levi-Civita connection if it is torsion-free, i.e., it satisfies ${\nabla}_{V}W-{\nabla}_{W}V = [V,W]$, for every $V,W\in \mathcal{X}(M)$ and it is compatible with the Riemannian metric $\mathbb{G}$, meaning it satisfies $V(\langle W,U \rangle_{\mathbb{G}}) = \langle {\nabla}_{V}W,U \rangle_{\mathbb{G}} + \langle W,{\nabla}_{V}U \rangle_{\mathbb{G}}$, for all $V,W,U \in \mathcal{X}(M)$. A Levi-Civita connection is uniquely determined by the following equation:
\begin{align} \label{Levi_Connection_prop} 
    &2 \langle U, {\nabla}_{V}W \rangle_{\mathbb{G}}
         + \langle [V,U],W \rangle_{\mathbb{G}} + \langle [W,U],V \rangle_{\mathbb{G}} + U \langle V,W\rangle_{\mathbb{G}}   \notag \\
    & = \langle [V,W],U \rangle_{\mathbb{G}} + V \langle W,U\rangle_{\mathbb{G}} +  W \langle U,V \rangle_{\mathbb{G}},
\end{align}
where $[\cdot,\cdot]$ denotes the Lie bracket, given by $[V,W](f) = V(W(f))-W(V(f))$. 
We denote the product manifold of $M$ and $N$ by $M \times N$. For each $(p,q) \in M \times N$, the subsets $M \times q = \left \{ (r,q)\in M \times N \mid r \in M\right \}$ and $p \times N = \left \{ (p,r)\in M \times N \mid r \in N\right \}$ are submanifolds of $M \times N$. 
%

\subsection{Matrix Lie groups}
Let $G$ denote a matrix Lie group and $\mathfrak{g}$ denote the Lie algebra of $G$. For any $A,B \in \mathfrak{g}$, the Lie bracket of $A$ and $B$ is defined as $[A,B] = AB - BA$. 
For each $g \in G$, the left translation $L_{g}:G \to G$ is defined as $L_{g}(h) = gh$ for every $h \in G$. Its tangent map, $\d L_{g}:T_{h}G \to T_{gh}G$, satisfies $\d L_{g}v = gv$ for all $v \in T_{h}G$. 
A vector field $V \in \mathcal{X}(G)$ is said to be left-invariant if for every $g,h \in G$, $V(gh) = \d L_{g}V(h)$. 
A Riemannian metric $\mathbb{G}$ is said to be left-invariant if, for each $g,h \in G$ and each $u,v \in T_{h}G$, $\langle u,v \rangle_{\mathbb{G}(h)} = \langle \d L_{g}(u),\d L_{g}(v) \rangle_{\mathbb{G}(gh)}$. A Riemannian metric $\mathbb{G}$ is bi-invariant if $\mathbb{G}$ is invariant under both left and right translations.  
Let $e^{(\cdot)}:\mathfrak{g} \to G$ denote the exponential map on $G$.
For each $A \in \mathfrak{g}$, the curve $\zeta: \mathbb{R} \to G$, given by $t \mapsto \zeta(t) = e^{At}$, is the unique one-parameter group such that $\zeta^{\prime}(0) = A$. Let $V_{A}\in \mathcal{X}(G)$ represent the left-invariant vector field corresponding to $A$ at identity. For any $g \in G$, the integral curve $\gamma: \mathbb{R} \to G$ of $V_{A}$ passing through $g$ is given by $\gamma(t) = ge^{At}$.

\subsection{Properties of $\mathrm{SO(3)}$}\label{Subsec:Prop_SO3}
The special orthogonal group $\mathrm{SO(3)}$ consists of all $3 \times 3$ orthogonal matrices with a determinant of 1. The Lie algebra of $\mathrm{SO(3)}$, denoted by $\mathfrak{so}(3)$, consists of all $3 \times 3$ skew-symmetric matrices. There exsits a natural isomorphism $\hat{\cdot}:\mathbb{R}^{3} \to \mathfrak{so}(3)$ between $\mathbb{R}^{3}$ and $\mathfrak{so}(3)$, such that for any $ \boldsymbol{v},\boldsymbol{w} \in \mathbb{R}^{3}$, $\hat{\boldsymbol{v}}\boldsymbol{w}=\boldsymbol{v} \times \boldsymbol{w}$. Let the inverse map of $\hat{\cdot}$ denote by $(\cdot)^{\vee}: \mathfrak{so}(3) \to \mathbb{R}^{3}$, we have $(\hat{\boldsymbol{v}})^{\vee} = \boldsymbol{v}$.
The exponential map on $\mathrm{SO(3)}$ is defined for any $\hat{\boldsymbol{v}} \in \mathfrak{so}(3)$ as $e^{\hat{\boldsymbol{v}}} = I + \frac{\sin\left \| \boldsymbol{v} \right \|}{\left \| \boldsymbol{v} \right \|} \hat{\boldsymbol{v}} + \frac{1- \cos\left \| \boldsymbol{v} \right \|}{\left \| \boldsymbol{v} \right \|^{2}}\hat{\boldsymbol{v}}^{2}$, when $\left \| \boldsymbol{v} \right \| \neq 0$, and $e^{\hat{\boldsymbol{v}}} = I$ otherwise. Let the inverse map of $e^{(\cdot)}$ denoted as $\log(\cdot)$.
In the paper~\cite{park1995distance}, a bi-invariant Riemannian metric $\mathbb{G}$ is introduced on $\mathrm{SO}(3)$, defined by the inner product $\langle \cdot, \cdot \rangle_{\mathbb{G}(R)} : T_{R}\mathrm{SO}(3) \times T_{R}\mathrm{SO}(3) \to \mathbb{R}$ at each $R \in \mathrm{SO}(3)$. For any two tangent vectors $K_{1}$, $K_{2} \in T_{R}\mathrm{SO}(3)$, the inner product is given by $ \langle K_{1},K_{2} \rangle_{\mathbb{G}(R)} = \boldsymbol{u}_{1}^{\top}\boldsymbol{u}_{2}$, where $\boldsymbol{u}_{1}=(R^{\top}K_{1})^{\vee}$ and $\boldsymbol{u}_{2}=(R^{\top}K_{2})^{\vee}$. The bi-invariant distance function induced by $\mathbb{G}$ is a map $d: \mathrm{SO(3)} \times \mathrm{SO(3)} \to \mathbb{R}_{\ge 0}$ defined such that for any $R_{1}$, $R_{2} \in \mathrm{SO(3)}$, it satisfies $d(R_{1},R_{2}) = \left \| \log(R_{1}^{\top}R_{2}) \right \|$. For more details about differential geometry and Lie groups, we refer to classical textbooks~\cite{do1992riemannian,sahin2017riemannian,bullo2019geometric,gallier2020differential}. 

\subsection{Reachability analysis on manifolds}

Given a manifold $M$ and a vector field $V\in \mathcal{X}(M)$, we consider the following differential equation:
\begin{align}\label{eq:diff}
    \dot{x} = V(x)
\end{align}
An integral curve of $V$ at $p$ is a curve $\gamma:\mathcal{T}_{p} \to M$ such that $\gamma(0) = p$ and it satisfies~\eqref{eq:diff}, for every $t \in \mathcal{T}_{p}\subset \mathbb{R}$. If $\mathcal{T}_{p}$ is the largest interval on which $\gamma$ is well-defined, then $\gamma$ at $p$ is maximal. The flow of a vector field $V\in \mathcal{X}(M)$ is the unique mapping $\psi: \mathcal{D} \times \mathcal{T} \to M$, where $\mathcal{D}$ is an open neighborhood around $p$, such that $t \mapsto \psi(p,t)$ represents the maximal integral curve of $V$ passing through $p$. 
If $\mathcal{U}\subseteq M$ satisfies $\sup{\mathcal{T}_{p}} = +\infty$ for each $p \in \mathcal{U}$, then we say $V$ is forward complete on $\mathcal{U}$. Given an initial set $\mathcal{I} \subset M$, the reachable set of the dynamical system~\eqref{eq:diff} at time $T \ge 0$ is defined as
\begin{equation*}
    \mathcal{R}_{V}(\mathcal{I},T) = \left \{ \psi(x_0, T) \mid x_0 \in \mathcal{I} \right \}.
\end{equation*}
The reachable tube of the dynamical system~\eqref{eq:diff} over the time interval $[0,T]$ is defined by
\begin{equation*}
    \mathcal{R}_{V}(\mathcal{I},[0,T])  = \bigcup_{t \in [0,T]} \mathcal{R}_{V}(\mathcal{I},t).
\end{equation*}

\subsection{Problem formulation}
We consider the attitude control system given by the following dynamical system on $\mathrm{SO(3)} \times \mathbb{R}^{3}$~\cite{bullo2019geometric,lee2010geometric}:
\begin{align}
\begin{split}
    \dot{R}&=R \hat{\boldsymbol{\omega}}, \\
    J\dot{\boldsymbol{\omega}}&=-\hat{\boldsymbol{\omega}} J\boldsymbol{\omega}+\boldsymbol{\tau}, \label{eq:attitude_dynamic2}
\end{split}
\end{align}
where $R \in \mathrm{SO(3)}$ is the rotation matrix, $\boldsymbol{\omega} \in \mathbb{R}^{3}$ is the body angular velocity, $J \in \mathbb{R}^{3 \times 3}$ is the inertia matrix, and $\boldsymbol{\tau}\in \mathbb{R}^3$ is the input to the system. 
We assume that the input $\boldsymbol{\tau}$ is predefined using a state-feedback law $\boldsymbol{\tau} = \boldsymbol{\tau}(R,\boldsymbol{\omega})$. We introduce the vector field $X \in \mathcal{X}(\mathrm{SO(3)} \times \mathbb{R}^{3})$ to represent the system dynamics as follows:
\begin{align}\label{eq:attitude_v_f}
    X(R,\boldsymbol{\omega}) &= \left ( X_{R}(R,\boldsymbol{\omega}), X_{\boldsymbol{\omega}}(R,\boldsymbol{\omega}) \right ) \notag \\
    &= (R \hat{\boldsymbol{\omega}},J^{-1}\left(-\hat{\boldsymbol{\omega}} J\boldsymbol{\omega}+\boldsymbol{\tau}\right)),
\end{align}
where $X_{R}:\mathrm{SO(3)} \times \mathbb{R}^{3} \to T\mathrm{SO(3)}$ and $X_{\boldsymbol{\omega}}: \mathrm{SO(3)} \times \mathbb{R}^{3} \to T\mathbb{R}^{3} \cong \mathbb{R}^{3}$. 
 Note that by fixing the second variable of $X_{R}$ and the first variable of $X_{\boldsymbol{\omega}}$, we obtain the vector fields $X_{R}(\cdot,\boldsymbol{\omega}) \in \mathcal{X}(\mathrm{SO(3)})$ and $X_{\boldsymbol{\omega}}(R,\cdot) \in \mathcal{X}(\mathbb{R}^{3})$, which are defined on the submanifolds $\mathrm{SO(3)}$ and $\mathbb{R}^{3}$, respectively.
%

In this paper, we focus on the problem of safety verification for the attitude control system~\eqref{eq:attitude_dynamic2}. We assume that the attitude dynamics~\eqref{eq:attitude_dynamic2} starts from some initial condition $\mathcal{I}\subseteq \mathrm{SO}(3)\times \mathbb{R}^3$ and we consider the unsafe region of the state-space as a subset $\mathcal{U}_{\mathrm{unsafe}}\subseteq \mathrm{SO}(3)\times \mathbb{R}^3$. Our goal is to ensure that the attitude dynamics~\eqref{eq:attitude_dynamic2} does not enter the unsafe region $\mathcal{U}_{\mathrm{unsafe}}$ over a given period of time. Using the notion of reachable sets, we can formulate this safety verification problem as follows. Given a time $T>0$, we want to show that, for every $t\in [0,T]$, 
\begin{align}\label{eq:avoid-problem}
    \mathcal{R}_{X}(\mathcal{I},t) \cap  \mathcal{U}_{\mathrm{unsafe}}=\emptyset.
\end{align}
Verifying condition~\eqref{eq:avoid-problem} requires computing reachable sets of the attitude dynamics~\eqref{eq:attitude_dynamic2}. However, in general, computing the exact reachable set of nonlinear systems is computationally challenging due to the nonlinearity of the dynamics and the complexity of the ambient space. Indeed, verifying that a point belongs to the reachable set of a nonlinear system is known to be undecidable~\cite{CM:90}. 
As a result, approximation methods are commonly employed to enhance tractability of reachability analysis. In formal safety verification, over-approximating reachable sets is essential to guarantee system safety.  

Simulation-based reachability methods have proven to be an effective solution to this problem in Euclidean space~\cite{girard2006verification,donze2007systematic,duggirala2013verification,maidens2014reachability,fan2017simulation}. These methods guarantee over-approximations of reachable sets by leveraging numerical simulations and computing distance bounds between neighboring trajectories.  
However, extending these methods to systems on manifolds introduces several challenges.
The first problem is how to select a suitable metric structure on the manifold so that the associated distance between trajectories can be efficiently computed.
Is it possible to use parameterized metrics on manifolds and select the optimal one to get tight over-approximations of reachable sets?
Furthermore, once we obtain the resulting over-approximations of reachable sets on manifolds, how can they be processed, visualized, and presented using existing Euclidean tools and software?

In this paper, we tackle these challenges by introducing a family of novel left-invariant Riemannian metrics on $\mathrm{SO}(3) \times \mathbb{R}^3$, parameterized by positive definite matrices. This parametrization enables the formulation of the search for optimal metrics as an SDP problem.
Using contraction theory on the Riemannian manifold $\mathrm{SO}(3) \times \mathbb{R}^3$, we provide upper bounds for the distance between trajectories of the attitude dynamics. 
Additionally, we explore the set representation on $\mathrm{SO}(3) \times \mathbb{R}^3$ and demonstrate how it can be mapped into Euclidean charts, enabling further analysis and visualization using classical Euclidean tools.


\section{Contraction Theory on $\mathrm{SO}(3) \times \mathbb{R}^3$}\label{Sec:Contrac_thm}
In this section, we first propose a novel metric structure on $\mathrm{SO}(3) \times \mathbb{R}^3$, based on a family of positive-definite matrices on $\mathbb{R}^{3\times 3}$ and a family of left-invariant Riemannian metrics on $\mathrm{SO}(3)$. For a dynamical system on a Riemannian manifold, we use contraction theory to upper bound the incremental distance between its trajectories. 
By applying this general result to our proposed metric structure on $\mathrm{SO}(3) \times \mathbb{R}^3$, we provide a necessary and sufficient characterization of the incremental distance between trajectories of the attitude dynamics. This characterization is formulated through a positive definite matrix inequality.

\subsection{Metrics on $ \mathrm{SO(3)} \times \mathbb{R}^3$}\label{Sec:Metric}
A product Riemannian metric on a product manifold can be constructed by combining the metrics defined on its individual submanifolds. 
In Euclidean contraction theory, it is common to construct contraction metrics using a family of positive definite matrices~\cite{tsukamoto2021contraction}. In this paper, we adopt a similar approach for constructing Riemannian metrics on the manifold $ \mathrm{SO(3)} \times \mathbb{R}^3$.
Our proposed family of Riemannian contraction metrics on \( \mathrm{SO(3)} \times \mathbb{R}^3 \) consists of separate metrics on \( \mathbb{R}^3 \) and \( \mathrm{SO}(3) \), each parameterized by a positive definite matrix.

\subsubsection{Metrics on $\mathbb{R}^3$}

Given a positive definite matrix $P \in \mathbb{R}^{3 \times 3}$, we define the $P$-metric $\mathbb{G}_{P}:\mathbb{R}^{3} \times \mathbb{R}^{3} \to \mathbb{R}$ on the tangent space $T_{\boldsymbol{x}}\mathbb{R}^{3} \cong \mathbb{R}^{3}$ by $\langle \boldsymbol{u}, \boldsymbol{v} \rangle_{\mathbb{G}_{P}} = \boldsymbol{u}^{\top}P \boldsymbol{v}$, for any $\boldsymbol{u}, \boldsymbol{v} \in \mathbb{R}^{3}$. The $P$-metric induces a $P$-norm defined by $\left \| \boldsymbol{u} \right \|_{P} = \sqrt{\boldsymbol{u}^{\top} A^{\top}A \boldsymbol{u}} = \left \| A\boldsymbol{u} \right \|$ with $P = A^{\top}A$. Since the geodesics in ($\mathbb{R}^3$,$\mathbb{G}_{P}$) are straight lines~\cite[Problem 13.10]{lee2012smooth}, the distance function with respect to $\mathbb{G}_{P}$ can be computed as $d_{P}(\boldsymbol{x}_{1}, \boldsymbol{x}_{2}) = \left \| \boldsymbol{x}_{1} - \boldsymbol{x}_{2} \right \|_{P}$.

\subsubsection{A family of left-invariant metrics on $\mathrm{SO(3)}$}\label{Sec:left_invariant_metric}
Inspired by the construction of $P$-metric on $\mathbb{R}^3$, we propose a new parameterized class of left-invariant metrics for $\mathrm{SO(3)}$.
\begin{defn}[Left-invariant metrics]\label{eq:GQ}
Let $K_{1},K_{2} \in T_{R}\mathrm{SO}(3)$ be any two tangent vectors at $R \in \mathrm{SO}(3) $. $\mathbb{G}_{Q}$ is a class of left-invariant Riemannian metrics on $\mathrm{SO}(3)$, associated with a positive definite matrix $Q \in \mathbb{R}^{3 \times 3}$, so that
    \begin{equation} \label{def:G_Q}
    \langle K_{1},K_{2} \rangle_{\mathbb{G}_{Q}(R)} = \boldsymbol{u}_{1}^{\top}Q\boldsymbol{u}_{2},
    \end{equation}
where $\boldsymbol{u}_{1}=(R^{\top}K_{1})^{\vee}$ and $\boldsymbol{u}_{2}=(R^{\top}K_{2})^{\vee}$.
\end{defn}

For any $R_{1},R_{2} \in \mathrm{SO}(3)$, the induced left-invariant distance function, denoted as $d_{Q}(R_{1},R_{2})$, can be defined as the length of the minimal geodesic connecting $R_{1}$ and $R_{2}$ in the metric space $( \mathrm{SO}(3), \mathbb{G}_{Q})$. 

One should note that the expression for the bi-invariant distance function on $\mathrm{SO}(3)$, equipped with a bi-invariant Riemannian metric, can be easily derived, as shown in Section~\ref{Subsec:Prop_SO3}. This derivation relies on the fact that, for the bi-invariant Riemannian metric, geodesics coincide with the one-parameter subgroups generated by left-invariant vector fields~\cite[Section 21.4]{gallier2020differential}.
However, when $\mathrm{SO}(3)$ is equipped with a left-invariant Riemannian metric, the geodesics do not necessarily coincide with the one-parameter subgroups generated by left-invariant vector fields~\cite[Section 21.3]{gallier2020differential}.
This distinction makes it difficult to derive closed-form expressions for both the geodesics and the distance function $d_{Q}(R_{1},R_{2})$ on $(\mathrm{SO}(3), \mathbb{G}_{Q})$.
However, this does not compromise the effectiveness of the contraction analysis on the product manifold \( \mathrm{SO}(3) \times \mathbb{R}^{3} \), as we will derive an equivalent form of metric balls on \( (\mathrm{SO}(3),\mathbb{G}_{Q}) \), facilitating the construction and analysis of the resulting over-approximated sets using Euclidean tools. 
This will be discussed in Section~\ref{Sec:Metric_Ball_Representation}.

\subsubsection{Product metrics on $\mathrm{SO(3)}\times \mathbb{R}^3$}
Following~\cite[Example 2.7]{do1992riemannian}, a product metric on $\mathrm{SO}(3) \times \mathbb{R}^3$ can be defined as follows.
\begin{defn}[Product metrics]\label{def:product_metrics}
    If $\mathrm{SO(3)}$ is equipped with $\mathbb{G}_{Q}$ and $\mathbb{R}^{3}$ is equipped with $\mathbb{G}_{P}$, then the product manifold $\mathrm{SO(3)}\times \mathbb{R}^3$ has a product Riemannian metric $\mathbb{G} = \mathbb{G}_{Q} \times \mathbb{G}_{P}$, such that, for any $(K_{1},\boldsymbol{v}_{1}),(K_{2},\boldsymbol{v}_{2}) \in T_{(R,\boldsymbol{\omega})}(\mathrm{SO(3)} \times \mathbb{R}^{3})$,
    \begin{align}
        \langle (K_{1},\boldsymbol{v}_{1}),(K_{2},\boldsymbol{v}_{2}) \rangle_{\mathbb{G}(R,\boldsymbol{\omega})} &= \langle K_{1},K_{2} \rangle_{\mathbb{G}_{Q}(R)} + \langle \boldsymbol{v}_{1}, \boldsymbol{v}_{2} \rangle_{\mathbb{G}_{P}} \notag \\
        & =\boldsymbol{u}_{1}^{\top}Q\boldsymbol{u}_{2} + \boldsymbol{v}_{1}^{\top}P\boldsymbol{v}_{2},
    \end{align}
    where $\boldsymbol{u}_{1}=(R^{\top}K_{1})^{\vee}$ and $\boldsymbol{u}_{2}=(R^{\top}K_{2})^{\vee}$.
\end{defn}
According to~\cite[Example 10.14]{boumal2023introduction}, the induced distance function on $(\mathrm{SO}(3) \times \mathbb{R}^3,\mathbb{G})$ satisfies
\begin{equation*}
    d((R,\boldsymbol{\omega}),(R_{0},\boldsymbol{\omega}_{0}))= \sqrt{d^{2}_{Q}(R,R_{0}) + d^{2}_{P}(\boldsymbol{\omega},\boldsymbol{\omega}_{0})}.
\end{equation*}

\subsection{Bounds on incremental trajectory distance}\label{Sec:contracting_vector_fields}
Contraction theory provides a necessary and sufficient condition for the incremental convergence of multiple nearby trajectories toward a single trajectory~\cite{tsukamoto2021contraction}.
For systems on Riemannian manifolds, the concepts of contracting vector fields can be used to establish exponentially convergent bounds on the distance between nearby trajectories. We build on the results presented in~\cite{simpson2014contraction, bullo2021contraction} for contracting vector fields on manifolds, extending them to address cases where bounds may either converge or diverge.

\begin{prop}[Bounds on trajectory distance]
    \label{thm:contracting_vec}
    Let $(M,\mathbb{G})$ be a Riemannian manifold, and let $V$ be a vector field on $M$. Denote by $\psi(q,t)$ the flow of $V$. Denote by $\mathcal{R}_{V}(\mathcal{I},[0,T])$ the reachable tube of $V$ initiating from $\mathcal{I}$ between the time interval $[0,T]$. If there exist a constant $c \in \mathbb{R}$ and connected sets $\mathcal{I}, \mathcal{U} \subset M$ such that
    \begin{enumerate}
        \item $\mathcal{I}$ is strongly convex,
        \item $\mathcal{R}_{V}(\mathcal{I},[0,T])\subset \mathcal{U}$,
        \item $V$ is forward complete on $\mathcal{U}$,
        \item $ \langle {\nabla}_{v}V,v \rangle_{\mathbb{G}(q)} \le c \langle v,v \rangle_{\mathbb{G}(q)}$, $\forall q \in \mathcal{U}, \forall v \in T_{q}M$,
    \end{enumerate}
    then for any $q_{1},q_{2} \in \mathcal{I}$, it holds that
    \begin{equation*}
        d(\psi(q_{1},t),\psi(q_{2},t)) \le e^{ct}d(q_{1},q_{2}), \quad \forall t \in [0,T].
    \end{equation*}
\end{prop}
\begin{pf}
    This can be proved using the same approach outlined in~\cite[Theorem 2.3]{simpson2014contraction}. Since the condition for infinitesimal contraction remains valid for any reachable points originating from $\mathcal{I}$, and the strong convexity of $\mathcal{I}$ ensures that it is 1-reachable, we can derive exponential bounds for trajectories starting from $\mathcal{I}$. The bounds here can either converge or diverge, as the same arguments apply for $c \in \mathbb{R}$.\hfill ~\qed
\end{pf}

In the remainder of this subsection, we demonstrate how the metric structure on $\mathrm{SO}(3) \times \mathbb{R}^3$, defined in Section~\ref{Sec:Metric}, can be used to upper bound the incremental trajectory distance of attitude control systems evolving on $\mathrm{SO}(3) \times \mathbb{R}^3$.

\begin{thm}\label{Thm:contraction_SO3}
    Consider a vector field $X \in \mathcal{X}(\mathrm{SO(3)}\times \mathbb{R}^3)$ and 
    suppose there exist a strongly convex set $\mathcal{I}$ and a connected set $\mathcal{U}$such that the reachable tube $\mathcal{R}_{X}(\mathcal{I},[0,T]) \subset \mathcal{U}$. Furthermore,suppose that there exist a product Riemannian metric $\mathbb{G} = \mathbb{G}_Q\times \mathbb{G}_P$ defined by positive definite matrices $Q,P \in \mathbb{R}^{3 \times 3}$, and a constant $c \in \mathbb{R}$ such that, for all $(R,\boldsymbol{\omega}) \in \mathcal{U}$, the following condition holds:
    \begin{align}\label{eq:contracting_vec_f}
        \begin{bmatrix}
            \hat{\boldsymbol{\omega}}Q - Q\hat{\boldsymbol{\omega}} - 2cQ  & Q + A^{\top}P\\
            Q + PA & B^{\top}P + PB - 2cP 
          \end{bmatrix} \preceq 0,
    \end{align}
    where the matrix functions $A: \mathrm{SO(3)} \times \mathbb{R}^3 \to \mathbb{R}^{3 \times 3}$ and $B: \mathrm{SO(3)} \times \mathbb{R}^3 \to \mathbb{R}^{3 \times 3}$ are defined as follows: for any $(R,\boldsymbol{\omega}) \in \mathrm{SO(3)}\times \mathbb{R}^3$ and $\boldsymbol{\alpha},\boldsymbol{\beta} \in \mathbb{R}^3$,
    \begin{align*}
    &A(R,\boldsymbol{\omega}) \boldsymbol{\alpha} = \lim_{t \to 0} \frac{X_{\boldsymbol{\omega}}(R + tR\hat{\boldsymbol{\alpha}},\boldsymbol{\omega}) - X_{\boldsymbol{\omega}}(R,\boldsymbol{\omega})}{t}, \\
    &B_{ij}(R,\boldsymbol{\omega})= \frac{\partial }{\partial \omega_{j}}\left (X^i_{\boldsymbol{\omega}}(R,\cdot) \right )(\boldsymbol{\omega}), \qquad \forall i,j \in \left \{ 1,2,3 \right \}.
    \end{align*}
    Then, for any $(R_{1},\boldsymbol{\omega}_{1}),(R_{2},\boldsymbol{\omega}_{2}) \in \mathcal{I}$ and $t \in [0,T]$,
    \begin{align*}    
        d(\psi(R_{1},\boldsymbol{\omega}_{1},t),\psi(R_{2},\boldsymbol{\omega}_{2},t))\le e^{ct}&d((R_{1},\boldsymbol{\omega}_{1}),(R_{2},\boldsymbol{\omega}_{2})),
    \end{align*}
    where $d$ is the distance on the Riemannian manifold $(\mathrm{SO}(3)\times \mathbb{R}^3,\mathbb{G})$. 
\end{thm}
\begin{pf}
Consider the product manifold $\mathrm{SO(3)} \times \mathbb{R}^3$, equipped with a product Riemannian metric defined as $\mathbb{G} = \mathbb{G}_{Q} \times \mathbb{G}_{P}$. Let $\nabla$ denote the Levi-Civita connection of $(\mathrm{SO(3)} \times \mathbb{R}^3, \mathbb{G})$. Following Proposition~\ref{thm:contracting_vec}, the condition for infinitesimal contraction is formulated as:
\begin{align}\label{ieq:orig_contraction}
    &\langle \nabla_{(K,v)}X,(K,v) \rangle_{\mathbb{G}(R,\boldsymbol{\omega})} \le c\langle (K,v), (K,v) \rangle_{\mathbb{G}{(R,\boldsymbol{\omega})}}, \notag \\
    & \quad \forall (R,\boldsymbol{\omega}) \in \mathcal{U}, \quad \forall (K,v) \in T_{(R,\boldsymbol{\omega})}(\mathrm{SO(3)} \times \mathbb{R}^{3}).
\end{align}
Let $\overset{R}{\nabla}$ and $\overset{\boldsymbol{\omega}}{\nabla}$ denote the Levi-Civita connections on the submanifolds $(\mathrm{SO(3)}, \mathbb{G}_{Q})$ and $(\mathbb{R}^3, \mathbb{G}_{P})$, respectively. Following~\cite[Exercise 5.4]{boumal2023introduction}, the covariant derivative of $X$ can be decomposed as
\begin{align*}
    \nabla_{(K,v)}X(R,\boldsymbol{\omega}) =& \left(  {\overset{R}{\nabla}}_{K}X_{R}(\cdot,\boldsymbol{\omega})(R) + \mathcal{D}_{v}X_{R}(R,\cdot)(\boldsymbol{\omega}), \right.\\
    & \left. {\overset{\boldsymbol{\omega}}{\nabla}_{v}X_{\boldsymbol{\omega}}}(R,\cdot)(\boldsymbol{\omega}) + \mathcal{D}_{K}X_{\boldsymbol{\omega}}(\cdot,\boldsymbol{\omega})(R) \right),
\end{align*}
where $\mathcal{D}_{v}X_{R}$ and $\mathcal{D}_{K}X_{\boldsymbol{\omega}}$ represent the differentials of the vector fields, given by
\begin{align*}
    \mathcal{D}_{v}X_{R}(R,\cdot)(\boldsymbol{\omega}) &= \lim_{t \to 0} \frac{X_{R}(R,\boldsymbol{\omega}+tv) - X_{R}(R,\boldsymbol{\omega})}{t}, \\
    \mathcal{D}_{K}X_{\boldsymbol{\omega}}(\cdot,\boldsymbol{\omega})(R) & = \lim_{t \to 0} \frac{X_{\boldsymbol{\omega}}(R + tK,\boldsymbol{\omega}) - X_{\boldsymbol{\omega}}(R,\boldsymbol{\omega})}{t}.
\end{align*}
By applying the definition of the product metric, one can easily obtain
\begin{align*}
    &\langle {\overset{R}{\nabla}}_{K}X_{R}(\cdot,\boldsymbol{\omega})(R) + \mathcal{D}_{v}X_{R}(R,\cdot)(\boldsymbol{\omega}) - c K, K \rangle_{\mathbb{G}_{Q}(R)} \\
    \le &- \langle {\overset{\boldsymbol{\omega}}{\nabla}_{v}X_{\boldsymbol{\omega}}}(R,\cdot)(\boldsymbol{\omega}) + \mathcal{D}_{K}X_{\boldsymbol{\omega}}(\cdot,\boldsymbol{\omega})(R) - cv, v \rangle_{\mathbb{G}_{P}}.
\end{align*} 
We can introduce bases for  
$\mathcal{X}(\mathrm{SO(3)})$ and $\mathcal{X}(\mathbb{R}^3)$ on the respective sides of the inequality and evaluate the condition within each space independently.

Let $\{ \frac{\partial }{\partial \omega_{1}}, \frac{\partial }{\partial \omega_{2}}, \frac{\partial }{\partial \omega_{3}}\}$ denote the standard basis of $\mathcal{X}(\mathbb{R}^3)$. It follows that $\overset{\boldsymbol{\omega}}{\nabla}_{\frac{\partial }{\partial \omega_{i}}}\frac{\partial }{\partial \omega_{j}} = 0$, for each $ i,j \in \left\{1,2,3\right\}$. The vector field $X_{\boldsymbol{\omega}}(R,\cdot)$ is then represented as $X_{\boldsymbol{\omega}}(R,\cdot)(\boldsymbol{\omega}) = \sum_{i=1}^{3} X^i_{\boldsymbol{\omega}} \frac{\partial }{\partial \omega_{i}}$. For any vector $v \in T\mathbb{R}^{3}\cong \mathbb{R}^{3}$, there exists a unique $\boldsymbol{\beta} \in \mathbb{R}^{3}$ such that $v = \sum_{i=1}^{3} \beta_{i} \frac{\partial }{\partial \omega_{i}}$. Applying~\eqref{Connection_prop_1} and~\eqref{Connection_prop_2},
the covariant derivative of $X_{\boldsymbol{\omega}}(R,\cdot)$ in this coordinate satisfies that
\begin{align}\label{eq:covar_deriv_R3}
    \overset{\boldsymbol{\omega}}{\nabla}_{v}X_{\boldsymbol{\omega}}(R,\cdot)(\boldsymbol{\omega}) & = \sum_{i,j=1}^{3} \beta_{i} \frac{\partial }{\partial \omega_{i}}(X^j_{\boldsymbol{\omega}})\frac{\partial }{\partial \omega_{j}} + \beta_{i}X^j_{\boldsymbol{\omega}}\overset{\boldsymbol{\omega}}{\nabla}_{\frac{\partial }{\partial \omega_{i}}} \frac{\partial }{\partial \omega_{j}} \notag \\
    & = \sum_{i,j=1}^{3} \beta_{i}  \frac{\partial }{\partial \omega_{i}}(X^j_{\boldsymbol{\omega}})\frac{\partial }{\partial \omega_{j}}.
\end{align}
Then, the right-hand side of the contraction condition can be reformulated as
\begin{align}\label{eq:right_contraction_ieq}
    & - \langle {\overset{\boldsymbol{\omega}}{\nabla}_{v}X_{\boldsymbol{\omega}}}(R,\cdot)(\boldsymbol{\omega}) + \mathcal{D}_{K}X_{\boldsymbol{\omega}}(\cdot,\boldsymbol{\omega})(R) - cv, v \rangle_{\mathbb{G}_{P}} \notag \\
    = & - \left \langle \mathcal{J}_{\boldsymbol{\beta}} + \mathcal{J}_{\boldsymbol{\alpha}} - c  \boldsymbol{\beta},  \boldsymbol{\beta} \right \rangle_{\mathbb{G}_{P}} \notag \\
    = & - (\mathcal{J}_{ \boldsymbol{\beta}} + \mathcal{J}_{\boldsymbol{\alpha}})^{\top}P \boldsymbol{\beta}+  c \boldsymbol{\beta}^{\top}P \boldsymbol{\beta},
\end{align}
where $\mathcal{J}_{ \boldsymbol{\beta}}, \mathcal{J}_{\boldsymbol{\alpha}} \in \mathbb{R}^{3}$ is defined as 
\begin{align}\label{eqn:Def_Ja_Jb}
\begin{split}
&\mathcal{J}_{\boldsymbol{\beta}i} = \sum_{j=1}^{3} \beta_{j}  \frac{\partial }{\partial \omega_{j}}\left (X^i_{\boldsymbol{\omega}}(R,\cdot) \right )(\boldsymbol{\omega}), \quad \forall i \in \left \{ 1,2,3 \right \},\\
&\mathcal{J}_{\boldsymbol{\alpha}i} = \lim_{t \to 0} \frac{X^i_{\boldsymbol{\omega}}(R + tK,\boldsymbol{\omega}) - X^i_{\boldsymbol{\omega}}(R,\boldsymbol{\omega})}{t}.
\end{split}
\end{align}
The first equality expresses the formula in the selected basis, using the definition of the differentials of the vector fields, while the second equality follows from the definition of the $P$-metric on $\mathbb{R}^3$.

Let $\{X_{L_{1}},X_{L_{2}},X_{L_{3}}\}$ represent a basis for $\mathcal{X}(\mathrm{SO(3)})$, where each basis element $X_{L_{i}}$ is defined by $X_{L_{i}}(R) = R L_{i} = R\hat{\boldsymbol{e}}_{i}$, for each $i \in \left \{ 1,2,3 \right \}$. Here, the vectors 
$\boldsymbol{e}_{1} = [1,0,0]^{\top},\boldsymbol{e}_{2} = [0,1,0]^{\top}, \boldsymbol{e}_{3} = [0,0,1]^{\top}$ form the standard basis of $\mathbb{R}^{3}$. 
The vector field $X_{R}(\cdot,\boldsymbol{\omega}) \in \mathcal{X}(\mathrm{SO(3)})$ is represented as $X_{R}(\cdot,\boldsymbol{\omega})(R) = \sum_{i=1}^{3} \omega_{i}X_{L_{i}}(R)$. For any vector $K \in T_{R}\mathrm{SO(3)}$, there exists a unique $\boldsymbol{\alpha} \in \mathbb{R}^{3}$ such that $K = \sum_{i=1}^{3} \alpha_{i}X_{L_{i}}(R) = R\hat{\boldsymbol{\alpha}}$. Applying~\eqref{Connection_prop_1} and~\eqref{Connection_prop_2} to the covariant derivative of $X_{R}(\cdot,\boldsymbol{\omega})$, we have
\begin{align}\label{eq:cov_deriv_XR}
    {\overset{R}{\nabla}}_{K}X_{R}(\cdot,\boldsymbol{\omega}) &= \sum_{i,j=1}^{3} \alpha_{i} \left ( X_{L_{i}}(\omega_{j})X_{L_{j}} + \omega_{j} {\overset{R}{\nabla}}_{X_{L_{i}}}X_{L_{j}} \right )  \notag\\
    & = \sum_{i,j=1}^{3} \alpha_{i} \omega_{j} {\overset{R}{\nabla}}_{X_{L_{i}}}X_{L_{j}}.
\end{align}
The second equality relies on the fact that the value of $\boldsymbol{\omega}$ is constant in this configuration.
For each $i,j \in \left \{ 1,2,3 \right \}$, the covariant derivative term ${\overset{R}{\nabla}}_{X_{L_{i}}}X_{L_{j}}$ can be viewed as a new vector field and expressed in terms of a coefficient vector $\boldsymbol{\varsigma}_{i,j}\in \mathbb{R}^{3}$, such that 
\begin{align}\label{eq:cov_deri_SO3basis1}
    {\overset{R}{\nabla}}_{X_{L_{i}}}X_{L_{j}}(R) = \sum_{k = 1}^{3} \varsigma^{k}_{i,j}X_{L_{k}}(R) = R \hat{\boldsymbol{\varsigma}}_{i,j}.
\end{align}
The coefficient vector $\boldsymbol{\varsigma}_{i,j}$ is uniquely determined by the Riemannian metric $\mathbb{G}_{Q}$ and satisfies the following equation:
\begin{align}\label{eq:cov_deri_SO3basis2}
    \sum_{i,j = 1}^{3}\alpha_{i}\omega_{j}\boldsymbol{\varsigma}_{i,j} = Q^{-1} \hat{\boldsymbol{\omega}}Q\boldsymbol{\alpha}-\frac{1}{2}\tr(Q) Q^{-1}\hat{\boldsymbol{\omega}}\boldsymbol{\alpha}
\end{align}
For the sake of brevity in this proof, we place the proof of Equation~\eqref{eq:cov_deri_SO3basis2} in the Appendix~\ref{Sec:App_Coeff_Vec}. By substituting Equations~\eqref{eq:cov_deriv_XR},~\eqref{eq:cov_deri_SO3basis1}, and~\eqref{eq:cov_deri_SO3basis2} into the left-hand side of the contraction condition, we obtain
\begin{align}\label{eq:left_contraction_ieq}
    &\langle {\overset{R}{\nabla}}_{K}X_{R}(\cdot,\boldsymbol{\omega})(R) + \mathcal{D}_{v}X_{R}(R,\cdot)(\boldsymbol{\omega}) - c K, K \rangle_{\mathbb{G}_{Q}(R)} \notag \\
    = & \langle \sum_{i,j = 1}^{3}\alpha_{i}\omega_{j}R\hat{\boldsymbol{\varsigma}}_{i,j} + R\hat{\boldsymbol{\beta}} - c R \hat{\boldsymbol{\alpha}} , R \hat{\boldsymbol{\alpha}} \rangle_{\mathbb{G}_{Q}(R)} \notag \\
    = & \sum_{i,j = 1}^{3}\alpha_{i}\omega_{j}\boldsymbol{\varsigma}_{i,j}^{\top} Q \boldsymbol{\alpha} + \boldsymbol{\beta}^{\top} Q \boldsymbol{\alpha} - c\boldsymbol{\alpha}^{\top} Q \boldsymbol{\alpha} \notag \\
    = & \boldsymbol{\alpha}^{\top} \left (\tr(Q)\hat{\boldsymbol{\omega}}/2 - Q\hat{\boldsymbol{\omega}} - cQ \right )\boldsymbol{\alpha} + \boldsymbol{\beta}^{\top} Q \boldsymbol{\alpha}.
\end{align}
The first equality expresses all the vectors in the basis of $\mathcal{X}(\mathrm{SO(3)})$ and applies the derivation $\mathcal{D}_{v}X_{R}(R,\cdot)(\boldsymbol{\omega}) = \lim_{t \to 0} (X_{R}(R,\boldsymbol{\omega}+tv) - X_{R}(R,\boldsymbol{\omega}))/t = R\hat{\boldsymbol{\beta}}$. The second equality follows from the definition of the Riemannian metric $\mathbb{G}_{Q}$ on $\mathrm{SO(3)}$, while the third equality utilizes the results from Equation~\eqref{eq:cov_deri_SO3basis2}.

By combining Equations~\eqref{eq:right_contraction_ieq} and~\eqref{eq:left_contraction_ieq}, we obtain the infinitesimal contraction condition
\begin{align}\label{ieq:contraction_step1}
    &\boldsymbol{\alpha}^{\top} \left ( \tr(Q)\hat{\boldsymbol{\omega}}/2 - Q\hat{\boldsymbol{\omega}} - cQ \right )\boldsymbol{\alpha} +  \boldsymbol{\beta}^{\top} Q \boldsymbol{\alpha} \notag \\
    & \qquad \qquad \qquad \le - (\mathcal{J}_{ \boldsymbol{\beta}} + \mathcal{J}_{\boldsymbol{\alpha}})^{\top}P \boldsymbol{\beta} +  c \boldsymbol{\beta}^{\top}P \boldsymbol{\beta}
\end{align}
which must hold for all $(R,\boldsymbol{\omega}) \in \mathcal{U}$ and $\boldsymbol{\alpha}, \boldsymbol{\beta} \in \mathbb{R}^{3}$. Observe that the expressions for $\mathcal{J}_{\boldsymbol{\beta}}$ and $\mathcal{J}_{\boldsymbol{\alpha}}$ in Equations~\eqref{eqn:Def_Ja_Jb} are essentially linear with respect to $\boldsymbol{\beta}$ and $\boldsymbol{\alpha}$, respectively. They can be reformulated as:
\begin{align*}
    \mathcal{J}_{\boldsymbol{\alpha}} = A(R,\boldsymbol{\omega}) \boldsymbol{\alpha}, \qquad \mathcal{J}_{\boldsymbol{\beta}} = B(R,\boldsymbol{\omega}) \boldsymbol{\beta},
\end{align*}
where the matrix function $A: \mathrm{SO(3)} \times \mathbb{R}^3 \to \mathbb{R}^{3 \times 3}$ and $B: \mathrm{SO(3)} \times \mathbb{R}^3 \to \mathbb{R}^{3 \times 3}$ can be determined for every $(R, \boldsymbol{\omega}) \in \mathrm{SO(3)} \times \mathbb{R}^3$ using Equation~\eqref{eqn:Def_Ja_Jb}. Then, the contraction condition~\eqref{ieq:contraction_step1} can be reformulated as:
\begin{align*}
    &\boldsymbol{x}^{\top}W\boldsymbol{x} \le 0 \qquad \forall \boldsymbol{x} = [\boldsymbol{\alpha};\boldsymbol{\beta}] \in \mathbb{R}^{6}, (R,\boldsymbol{\omega}) \in \mathcal{U},\\
    &W = \begin{bmatrix}
        \tr(Q)\hat{\boldsymbol{\omega}}/2 - Q\hat{\boldsymbol{\omega}} - cQ  & A^{\top}P\\
        Q & B^{\top}P - cP 
      \end{bmatrix}.
\end{align*}
This is equivalent to $W+W^{T}\preceq 0$, for all $(R,\boldsymbol{\omega}) \in \mathcal{U}$, resulting in the following condition:
\begin{align*}
    \begin{bmatrix}
        \hat{\boldsymbol{\omega}}Q - Q\hat{\boldsymbol{\omega}} - 2cQ  & Q + A^{\top}P\\
        Q + PA & B^{\top}P + PB - 2cP 
      \end{bmatrix} \preceq 0,
\end{align*}
which must hold for all $(R,\boldsymbol{\omega}) \in \mathcal{U}$.\hfill ~\qed
\end{pf}


\section{Simulation-based Geometric Reachability}\label{Sec:Reach_methods}

In this section, we combine the incremental trajectory bounds from Theorem~\ref{Thm:contraction_SO3} with simulations of the attitude dynamics~\eqref{eq:attitude_dynamic2} to obtain an over-approximation of the system's reachable sets. Our contraction-based approach provides flexibility in selecting both the metric, known as the contraction metric, and the contraction rate $c$. In the following section, we will show how these over-approximations can be formulated as an SDP problem and efficiently solved within the geometric reachability framework.
Given an initial set $\mathcal{I}\subset \mathrm{SO}(3) \times \mathbb{R}^3$, our goal in this section is to over-approximate the reachable set $\mathcal{R}_{X}(\mathcal{I},T)$ using a finite number of simulations. We denote the metric balls in $(\mathbb{R}^3, \mathbb{G}_{P} )$ by $B_{P}$, the metric balls in $(\mathrm{SO}(3), \mathbb{G}_{Q})$ by $\mathcal{B}_{Q}$ and the metric balls in the product manifold $(\mathrm{SO}(3) \times \mathbb{R}^3, \mathbb{G}_{Q} \times \mathbb{G}_{P})$  by $\mathcal{B}_{Q \times P}$.

\subsection{Decomposition and partition of initial sets}

For a given initial set $\mathcal{I} \subset \mathrm{SO}(3) \times \mathbb{R}^3$, we first decompose it into two minimal independent sets $\mathcal{I}_{R} \subset \mathrm{SO(3)}$ and $\mathcal{I}_{V} \subset \mathbb{R}^{3}$, such that their Cartesian product satisfies $\mathcal{I} \subset \mathcal{I}_{R} \times \mathcal{I}_{V}$. 
In $\mathrm{SO}(3)$, the rotation set $\mathcal{I}_{R}$ can be covered by a finite number of standard balls $\mathcal{B}_{I}$ with the same radius. This is achieved by generating uniform deterministic samples on $\mathrm{SO(3)}$. Several methods have been proposed for this, including the successive orthogonal images~\cite{tan2020constrained,mitchell2008sampling,yershova2010generating} and the layered Sukharev approach~\cite{yershova2004deterministic,lindemann2005incremental}. Since these sampling techniques ensure global coverage and separation on $\mathrm{SO}(3)$, we can define standard balls $\mathcal{B}_{I}$ centered at the sample points, with a uniform radius determined by the sampling step. 
Since the bi-invariant distance function is explicit, identifying a collection of balls that completely covers the set $\mathcal{I}_{R}$ becomes straightforward. Covering the angular velocity set $\mathcal{I}_{V}$ is simpler, as it can always be covered by a collection of ellipsoids $B_{P}$ in $\mathbb{R}^{3}$. Now, the initial set is covered by a union of product metric balls, denoted as $\mathcal{B}_{I \times P}$, for each pair $(\mathcal{B}_{I}, B_{P})$ in the collections of $\mathcal{B}_{I}$ and $B_{P}$. The computation of the reachable set for each pair can be performed in parallel. By partitioning the space into finer segments with smaller step sizes, a more accurate approximation of the overall reachable set can be achieved.

Note that any metric ball in a Riemannian manifold with a sufficiently small radius is strongly convex~\cite[Theorem 5.3]{sakai1996riemannian}. 
We assume that every initial set has been processed and partitioned such that the radius of every initial ball is small enough to guarantee the validity of the strong convexity arguments in Theorem~\ref{Thm:contraction_SO3} throughout the reachable set computation process. 
For simplicity, the initial set in the proposed algorithm is provided as
$\mathcal{I} = \mathcal{B}_{I \times P_{0}}((R_{0},\boldsymbol{\omega}_{0}),r_{0})$.

\subsection{Contraction-based reachability (ConReach) algorithm}
In this section, we introduce the Contraction-based Reachability (ConReach) algorithm, which computes an over-approximation \( \overline{\mathcal{R}}_{X} \) of the reachable set \( \mathcal{R}_{X} \) for the attitude dynamics~\eqref{eq:attitude_dynamic2}. The key idea is to over-approximate the reachable set by expanding around a finite number of simulations, where the expansion factor is determined by the optimal incremental trajectory bound derived from contraction theory.

The time interval $[0,T]$ is divided into smaller segments, denoted by $[t_{i},t_{i+1}]$, where $i \in \left\{ 0, 1, \ldots, N-1 \right\}$, $t_{0} = 0$ and $t_{N} = T$.
This segmentation enables the computation of an optimal bound for each time interval. 
Let $\left \{ (R_{i},\boldsymbol{\omega}_{i}) \right \}$ denote a sequence of time-stamped states along the nominal simulation initiated from $(R_{0},\boldsymbol{\omega}_{0})$, where $(R_{i},\boldsymbol{\omega}_{i}) = \psi(R_{0},\boldsymbol{\omega}_{0},t_{i})$. 
At each step $i$, the over-approximated reachable set $\overline{\mathcal{R}}_{X}(\mathcal{I},t_{i+1})$ is calculated for the terminal time $t_{i+1}$. Let $\mathcal{B}_{Q_{i} \times P_{i}}((R_{i},\boldsymbol{\omega}_{i}), r_{i})$ represent the over-approximated reachable set at $t_{i}$, which serves as the terminal reachable set for the interval $[t_{i-1},t_{i}]$ at step $i-1$. During the time interval $[t_{i},t_{i+1}]$, the goal is to determine the optimal matrices $Q_{i+1},P_{i+1}$ and contraction rate $c_{i+1}$ that minimize the size of the over-approximated region at $t_{i+1}$. We introduce a line search over the value of contraction rate $c_{i+1}$. For a fixed $c_{i+1}$, we establish an SDP problem searching for the optimal matrices $Q_{i+1}$ and $P_{i+1}$.
In this paper, the optimization objective is defined as $f(Q_{i+1},P_{i+1}) = -\tr(Q_{i+1})$, which corresponds to minimizing the projected volume of the initial set on $\mathrm{SO}(3)$ at $t_{i}$.
%
The first constraint ensures that the initial set at step $i$, represented as metric balls with respect to $\mathbb{G}_{Q_{i+1} \times P_{i+1}}$, sufficiently covers the terminal set $\mathcal{B}_{Q_{i} \times P_{i}}((R_{i},\boldsymbol{\omega}_{i}), r_{i})$ at step $i-1$. 
While the explicit mathematical expression for these metric balls may not always be available, the next lemma provides a sufficient condition to establish the inclusion relationship between the metric balls in different metric spaces.
\begin{lem}[Norm ball inclusion]\label{lem:inclusion_product}
   Consider two metric balls $\mathcal{B}_{Q_{1} \times P_{1}}((R_{0},\boldsymbol{\omega}_{0}),r)$ and $\mathcal{B}_{Q_{2} \times P_{2}}((R_{0},\boldsymbol{\omega}_{0}),r)$ on the manifold $\mathrm{SO}(3)\times \mathbb{R}^{3}$ both centered at the configuration $(R_{0},\boldsymbol{\omega}_{0})$ with radius $r \in \mathbb{R}_{\ge 0}$. If $Q_{1} \succeq Q_{2}$ and $P_{1} \succeq P_{2}$, then $\mathcal{B}_{Q_{1} \times P_{1}}((R_{0},\boldsymbol{\omega}_{0}),r) \subseteq \mathcal{B}_{Q_{2} \times P_{2}}((R_{0},\boldsymbol{\omega}_{0}),r)$. 
\end{lem}
\begin{pf}
    The proof is attached in the Appendix~\ref{Sec:App_Inclu_Relation}.\hfill ~\qed
\end{pf}

\begin{figure*}[ht]
    \centering
    \begin{equation}\label{prob:optimization_local}
        \begin{aligned}
            \min \quad &  -\tr(Q_{i+1})\\
            \textrm{s.t.} 
            \quad & Q_{i+1} \succeq 0, \quad Q_{i+1}\preceq Q_{i}, \quad P_{i+1} \succeq 0, \quad P_{i+1}\preceq P_{i},\\
            \quad & \begin{bmatrix}
                \hat{\boldsymbol{\omega}}Q_{i+1} - Q_{i+1}\hat{\boldsymbol{\omega}} - 2c_{i+1}Q_{i+1}  & Q_{i+1} + A^{\top}P_{i+1}\\
                Q_{i+1} + P_{i+1}A & B^{\top}P_{i+1} + P_{i+1}B - 2c_{i+1}P_{i+1} 
              \end{bmatrix} \preceq 0, \quad \forall A \in \mathcal{A}_{i}, B \in \mathcal{B}_{i}, \hat{\boldsymbol{\omega}} \in \mathcal{O}_{i}, \\
        \end{aligned}
    \end{equation}
\end{figure*}

The second constraint arises from the matrix inequality condition~\eqref{eq:contracting_vec_f}. Inspired by the work of Fan~\cite{fan2017simulation}, we introduce interval arithmetic in this constraint.
Let $\mathcal{A}_{i}$, $\mathcal{B}_{i}$ and $\mathcal{O}_{i}$ denote the interval matrices such that, for all $(R, \boldsymbol{\omega}) \in \mathcal{U}_{i}$, $A(R, \boldsymbol{\omega}) \in \mathcal{A}_{i}$, $B(R, \boldsymbol{\omega}) \in \mathcal{B}_{i}$ and $\hat{\boldsymbol{\omega}} \in \mathcal{O}_{i}$, respectively. The set $\mathcal{U}_{i}$ can be computed from a coarse over-approximation of the reachable tube $\mathcal{R}_{X}(\mathcal{I}, [t_i, t_{i+1}])$ based on user information or using Riemannian Lipschitz bounds~\cite{shafa2024guaranteed}, and can be further refined through iterative optimization of prior estimations.
The SDP problem is then formulated in Problem~\eqref{prob:optimization_local}. In~\cite{fan2017simulation}, the authors present several methods for solving these SDP problems using interval arithmetic. However, we will not delve into this process here and will instead focus on the resulting solution.

By bloating the distance bounds of the states obtained from the time-stamped simulation at $t_{i+1}$, we construct a metric ball centered at $(R_{i+1}, \boldsymbol{\omega}_{i+1})$, which contains all the reachable set at $t_{i+1}$:
\begin{align*}
    \overline{\mathcal{R}}_{X}(\mathcal{I},t_{i+1}) &= \mathcal{B}_{Q_{i+1} \times P_{i+1}}((R_{i+1},\boldsymbol{\omega}_{i+1}), r_{i+1}), \\
    r_{i+1} &= r_{i}e^{c_{i+1}(t_{i+1}-t_{i})}.
\end{align*}
The ConReach algorithm is provided as Algorithm~\ref{alg:main}.
\begin{algorithm}[ht]
    \caption{ConReach}\label{alg:main}
     \hspace*{\algorithmicindent} \textbf{Input:} An initial set $\mathcal{I} = \mathcal{B}_{I \times P_{0}}((R_{0},\boldsymbol{\omega}_{0}),r_{0})$, time-stamped states $\left \{ (R_{i},\boldsymbol{\omega}_{i}),t_{i} \right \} $, the total number of time-stamped states $N$, terminal time $T$, the number of line search steps $N_s$, the range of the line search $[c_{\min},c_{\max}]$.\\
     \hspace*{\algorithmicindent} \textbf{Output:} $\overline{\mathcal{R}}_{X}(\mathcal{I},T)$ 
    \begin{algorithmic}[1] 
    \State $P_{0} \gets P_{0}$, $Q_{0} \gets I$;
    \For {$i=0,1,2,\ldots,N-1$}
        \State $\mathcal{U}_{i} \gets $ GetReachTube$(\overline{\mathcal{R}}_{X}(\mathcal{I},t_{i}),[t_{i},t_{i+1}])$;
        \State $\mathcal{A}_{i},\mathcal{B}_{i},\mathcal{O}_{i} \gets $ GetIntervalMat$(\mathcal{U}_{i})$;
        \State $has\_sol = false$;
        \For {$j=0,1,2,\ldots,N_s$}
            \State $c_{tmp} \gets c_{\max} - j*(c_{\max}-c_{\min})/N_s$;
            \State $has\_sol,Q_{tmp},P_{tmp} \gets $ SolveOpt$(Q_{i},P_{i},\cdots$
            \State \qquad \qquad \qquad \qquad \qquad \quad  $\mathcal{A}_{i},\mathcal{B}_{i},\mathcal{O}_{i},c_{tmp})$;
            \If {$has\_sol$}
                \State $Q_{i+1} \gets Q_{tmp},P_{i+1} \gets P_{tmp}$;
                \State $c_{i+1} \gets c_{tmp}$;
            \Else 
                \State break;
            \EndIf
        \EndFor
        \State $r_{i+1} \gets r_{i}e^{c_{i+1}(t_{i+1}-t_{i})}$;
        \State $\overline{\mathcal{R}}_{X}(\mathcal{I},t_{i+1}) \gets \mathcal{B}_{Q_{i+1} \times P_{i+1}}((R_{i+1},\boldsymbol{\omega}_{i+1}), r_{i+1})$ 
    \EndFor
    \State \Return $\overline{\mathcal{R}}_{X}(\mathcal{I},T) \gets \mathcal{B}_{Q_{N} \times P_{N}}((R_{N},\boldsymbol{\omega}_{N}), r_{N})$
    \end{algorithmic}
\end{algorithm}

\section{Representation of Metric Balls}\label{Sec:Metric_Ball_Representation}

With the over-approximation sets now represented as metric balls on $\mathrm{SO(3)} \times \mathbb{R}^3$, a key question arises: how can these results be used for further analysis? Since most existing tools and software are designed for Euclidean problems, it is important to represent our results in Euclidean space, enabling general solvers and analytical methods to engage with this topic effectively.  
In this section, we introduce a practical approach to representing metric balls separately within $\mathbb{R}^{3}$ and the atlas of $\mathrm{SO}(3)$. This decomposition allows for the independent analysis of specifications related to the attitude and angular velocity of the attitude dynamics~\eqref{eq:attitude_dynamic2}.

We consider a metric ball $\mathcal{B}_{Q \times P}((R_{0},\boldsymbol{\omega}_{0}), r)$ on $\mathrm{SO}(3) \times \mathbb{R}^{3}$. Let $\pi: \mathrm{SO}(3) \times \mathbb{R}^{3} \to \mathrm{SO(3)}$ and $\rho: \mathrm{SO}(3) \times \mathbb{R}^{3} \to \mathbb{R}^{3}$ denote the canonical projection onto $\mathrm{SO(3)}$ and $\mathbb{R}^{3}$, respectively. It holds that 
\begin{align*}
    \pi(\mathcal{B}_{Q \times P}((R_{0},\boldsymbol{\omega}_{0}), r)) &= \mathcal{B}_{Q}(R_{0}, r), \\
    \rho(\mathcal{B}_{Q \times P}((R_{0},\boldsymbol{\omega}_{0}), r)) &= B_{P}(\boldsymbol{\omega}_{0}, r).
\end{align*}
Specifications in the angular velocity space can be handled directly, as the projected set of angular velocity lies within Euclidean space. However, for the projected rotation set, we may not have an explicit expression, but we provide an equivalent form that allows for a sufficiently precise representation in Euclidean space.

\subsection{Equivalent representation of metric balls on $\mathrm{SO}(3)$}

As discussed in Section~\ref{Sec:left_invariant_metric}, deriving the exact mathematical expression for geodesics on left-invariant Lie groups may not be feasible. However, their evolutions can still be effectively captured using left Lie reduction~\cite{holm2009geometric,zacur2014left,noakes2022finding}. We use the theory of left Lie reduction to derive the dynamics of the geodesics of $(\mathrm{SO}(3),\mathbb{G}_{Q})$. We start by introducing the notion of reduced curves on $\mathrm{SO}(3)$. 
\begin{defn}[Reduced curve]
    Given a smooth curve $\gamma: \mathbb{R} \to \mathrm{SO}(3)$, the reduced curve of $\gamma$, denoted as $\boldsymbol{w}: \mathbb{R} \to \mathbb{R}^{3}$, is defined as $\boldsymbol{w}(t)= (\gamma^{\top}(t)\dot{\gamma}(t))^{\vee}$, for every $t\in \mathbb{R}$.
\end{defn}

\begin{lem}[Reduced geodesic dynamics]\label{prop:geo_dynamics}
    Consider the Riemannian manifold $\mathrm{SO}(3)$ equipped with a left-invariant Riemannian metric $\mathbb{G}_{Q}$. 
    For any geodesic $\gamma_{R}: [0,1] \to \mathrm{SO}(3)$, its reduced curve $\boldsymbol{w}_{R}:[0,1] \to \mathbb{R}^{3}$ satisfies the following equation:
    \begin{equation}\label{eq:rgd}
        Q\dot{\boldsymbol{w}}_{R} = -\boldsymbol{w}_{R} \times Q\boldsymbol{w}_{R}.
    \end{equation}
\end{lem}
\begin{pf}
Since $\mathrm{SO(3)}$ is complete, by Hopf-Rinow theorem, any two points $R_{1}$ and $R_{2}$ can be joined by a unique geodesic of minimal length.
For all the curves $\gamma:[0,1] \to \mathrm{SO(3)}$ connecting $R_{1}$ and $R_{2}$, the geodesic is the extremal curve of the following optimization problem:
\begin{equation}
    \begin{aligned}
        \min_{\gamma} \quad & \frac{1}{2}\int_{0}^{1} \langle \dot{\gamma}(t),\dot{\gamma}(t)\rangle_{\mathbb{G}_{Q}(\gamma(t))} \d t\\
        \textrm{s.t.} \quad & \gamma(0)=R_{1},\gamma(1)=R_{2}  
    \end{aligned}
\end{equation}
Define $\hat{\boldsymbol{w}}(t) = \gamma^{\top}(t)\dot{\gamma}(t) \in \mathfrak{so}(3)$. By applying left Lie reduction~\cite[Section 3.3]{zacur2014left}, the above problem is equivalent to
\begin{equation}\label{eq:opt2}
    \begin{aligned}
        \min_{\hat{\boldsymbol{w}}(t)} \quad & \frac{1}{2}\int_{0}^{1} l(\hat{\boldsymbol{w}}(t)) \d t\\
        \textrm{s.t.} \quad & \gamma_{R}(0)=R_{1},\gamma_{R}(1)=R_{2}  
    \end{aligned}
\end{equation}
where $l : \mathfrak{so}(3) \to \mathbb{R}$ is a reduced Lagrangian defined as $l(\hat{\boldsymbol{v}}) = \boldsymbol{v} ^{\top} Q \boldsymbol{v}$ for every $\hat{\boldsymbol{v}} \in \mathfrak{so}(3)$. The optimal solution of~\eqref{eq:opt2} satisfies the variational equation $\delta \int_{0}^{1} l(\hat{\boldsymbol{w}}) \d t = \boldsymbol{0}$. Let $\delta \gamma_{R}$ be the variations of paths between two fixed points. We have $\delta \gamma_{R}(0) = \delta \gamma_{R}(1) = \boldsymbol{0}$. Then the induced variations $\delta \hat{\boldsymbol{w}}$ satisfies
\begin{align*}
    \delta \hat{\boldsymbol{w}} & = \delta(\gamma_{R}^{-1}\dot{\gamma}_{R}) = - \gamma_{R}^{-1}\delta\gamma_{R}\gamma_{R}^{-1}\dot{\gamma}_{R} + \gamma_{R}^{-1}\delta\dot{\gamma}_{R}\\
    & = - \hat{\boldsymbol{\varsigma}}\hat{\boldsymbol{w}} + \gamma_{R}^{-1}\delta\dot{\gamma}_{R},
\end{align*}
where $\hat{\boldsymbol{\varsigma}}(t) = \gamma_{R}^{-1}\delta\gamma_{R} \in \mathfrak{so}(3)$. One can easily check that $\hat{\boldsymbol{\varsigma}}(0) = \hat{\boldsymbol{\varsigma}} (1) = \boldsymbol{0}$. By differentiating $\hat{\boldsymbol{\varsigma}}$, we have
\begin{align*}
    \frac{\d}{\d t}\hat{\boldsymbol{\varsigma}} &=  - \gamma_{R}^{-1}\dot{\gamma}_{R}\gamma_{R}^{-1}\delta\gamma_{R} + \gamma_{R}^{-1}\delta\dot{\gamma}_{R} = - \hat{\boldsymbol{w}} \hat{\boldsymbol{\varsigma}} + \gamma_{R}^{-1}\delta\dot{\gamma}_{R}.
\end{align*}
Then it holds that 
\begin{align*}
    \delta \hat{\boldsymbol{w}} = - \hat{\boldsymbol{\varsigma}}\hat{\boldsymbol{w}} + \hat{\boldsymbol{w}} \hat{\boldsymbol{\varsigma}} + \frac{\d}{\d t}\hat{\boldsymbol{\varsigma}} = [\hat{\boldsymbol{w}},\hat{\boldsymbol{\varsigma}}] + \frac{\d}{\d t}\hat{\boldsymbol{\varsigma}}.
\end{align*}
Let $ \boldsymbol{w} = (\hat{\boldsymbol{w}})^{\vee}$.
We get the equivalent relation in the isomorphic space:
\begin{equation*}
    \delta \boldsymbol{w} = \boldsymbol{w} \times \boldsymbol{\varsigma} + \dot{\boldsymbol{\varsigma}}.
\end{equation*}
According to the variational principle, the extremal curve of $\boldsymbol{w}(t)$, denoted by $\boldsymbol{w}_{R}(t)$, satisfies that 
\begin{align*}
    \boldsymbol{0} & = \delta \int_{0}^{1} l(\hat{\boldsymbol{w}}) \d t = \int_{0}^{1} \langle \frac{\delta l}{\delta \boldsymbol{w}},\delta \boldsymbol{w} \rangle \d t \\
    & = \int_{0}^{1} \langle Q\boldsymbol{w} ,\boldsymbol{w} \times \boldsymbol{\varsigma} + \dot{\boldsymbol{\varsigma}} \rangle \d t \\
    & = \int_{0}^{1} \langle Q\boldsymbol{w} ,\boldsymbol{w} \times \boldsymbol{\varsigma} \rangle + \frac{\d}{\d t} \langle Q\boldsymbol{w} ,\boldsymbol{\varsigma} \rangle - \langle Q\dot{\boldsymbol{w}} ,\boldsymbol{\varsigma} \rangle \d t \\
    & = \int_{0}^{1} - \langle \boldsymbol{w} \times Q\boldsymbol{w} + Q\dot{\boldsymbol{w}} ,\boldsymbol{\varsigma} \rangle \d t + \int_{0}^{1} \frac{\d}{\d t} \langle Q\boldsymbol{w} ,\boldsymbol{\varsigma} \rangle \d t \\
    & = \int_{0}^{1} - \langle \boldsymbol{w} \times Q\boldsymbol{w} + Q\dot{\boldsymbol{w}} ,\boldsymbol{\varsigma} \rangle \d t
\end{align*}
Therefore, we have 
\begin{equation*}
     Q\dot{\boldsymbol{w}}_{R} = - \boldsymbol{w}_{R} \times Q\boldsymbol{w}_{R}
\end{equation*}
holds for any curve $\hat{\boldsymbol{\varsigma}} \in \mathfrak{so}(3)$. This describes the dynamics of the reduced geodesic $\boldsymbol{w}_{R}(t)$ on $\mathbb{R}^{3}$.\hfill ~\qed
\end{pf}

Using the reduced geodesic dynamics in~\eqref{eq:rgd}, we propose an equivalent representation of the metric balls in $(\mathrm{SO}(3),\mathbb{G}_{Q})$. 

\begin{prop}[Equivalent representation]\label{Prop:SO3_Ball}
    Define a geodesic system described by
    \begin{align}
        \begin{split}
        \dot{R}&=R \hat{\boldsymbol{w}} \\
        Q\dot{\boldsymbol{w}}& =-\hat{\boldsymbol{w}}Q\boldsymbol{w}.\label{eq:geo_dynamic}
    \end{split}
    \end{align}
    Denote by $\mathcal{R}_{G}(\mathcal{I}_{G},T)$ the reachable set of this system at time $T$, starting from $\mathcal{I}_{G} = R_{0} \times B_{Q}(\boldsymbol{0},r)$. It holds that
    \begin{equation}\label{eq:Eq_metric_ball}
        \mathcal{B}_{Q}(R_{0},r) = \pi(\mathcal{R}_{G}(\mathcal{I}_{G},1)).
    \end{equation}
\end{prop}
\begin{pf} 
    Equations~\eqref{eq:geo_dynamic} define the dynamics of the reduced geodesic, as proven in Lemma~\ref{prop:geo_dynamics}. Therefore, the set $\pi(\mathcal{R}_{G}(\mathcal{I}_{G},1))$ includes all possible rotations that the geodesic, starting from $\mathcal{I}_{G}$, can reach at $t = 1$. 
    Let the geodesic passing through $R \in \mathrm{SO}(3)$ with the initial velocity $K \in T_{R}\mathrm{SO}(3)$ denoted as $\gamma_{(R,K)}: \mathbb{R} \to \mathrm{SO}(3)$. We have $\gamma_{(R,K)}(0) = R$ and $\dot{\gamma}_{(R,K)}(0) = K$.

    First, consider any point $R_{1} \in \mathcal{B}_{Q}(R_{0},r)$. According to the Hopf-Rinow theorem, there exists a unique geodesic of minimal length connecting $R_{0}$ to $R_{1}$. Let this geodesic be denoted as $\gamma_{(R_{0},K)}(t)$, which reaches $R_{1}$ at $t = t_{1}$. By the homogeneity of geodesics~\cite[Lemma 2.6]{do1992riemannian}, this is equivalent to a geodesic represented by $\gamma_{(R_{0},Kt_{1})}$ that reaches $R_{1}$ at $t = 1$, meaning $\gamma_{(R_{0},K)}(t_{1}) = \gamma_{(R_{0},Kt_{1})}(1)$. The distance between $R_{0}$ and $R_{1}$ satisfies that
    \begin{align*}
        &d_{Q}(R_{0},R_{1}) = d_{Q}(R_{0},\gamma_{(R_{0},Kt_{1})}(1))\\
        = & \int_{0}^{1} \sqrt{\langle \dot{\gamma}_{(R_{0},Kt_{1})}(t),\dot{\gamma}_{(R_{0},Kt_{1})}(t) \rangle _{\mathbb{G}_{Q}(\gamma_{(R_{0},Kt_{1})}(t))} } \d t\\
        = & \int_{0}^{1} \sqrt{\langle Kt_{1},Kt_{1} \rangle_{\mathbb{G}_{Q}(R_{0})} } \d t = \sqrt{\boldsymbol{w}^{\top}_{1}Q\boldsymbol{w}_{1}},
    \end{align*}
    where $\boldsymbol{w}_{1} = (R_{0}^{-1}Kt_{1})^{\vee}$. The third equality above follows from the property that the arc length of geodesics remains constant~\cite[Definition 2.1]{do1992riemannian}. 
    Given that $R_{1} \in \mathcal{B}_{Q}(R_{0},r)$, it follows that $\sqrt{\boldsymbol{w}^{\top}_{1}Q\boldsymbol{w}_{1}} < r$, which is equivalent to $\boldsymbol{w}_{1} \in B_{Q}(\boldsymbol{0},r)$. Therefore, for any $R_{1} \in \mathcal{B}_{Q}(R_{0},r)$, it can be reached by $\gamma_{(R_{0},R_{0}\hat{\boldsymbol{w}}_{1})}(t)$, at time $t = 1$, leading to $\mathcal{B}_{Q}(R_{0},r) \subset \pi(\mathcal{R}_{G}(\mathcal{I}_{G},1))$.
    
    Next, consider any initial state $(R_{0},\boldsymbol{w}_{0}) \in \mathcal{I}_{G}$. Let the rotation reached by the geodesic $\gamma_{(R_{0},R_{0}\hat{\boldsymbol{w}}_{0})}(t)$ at $t = 1$ denoted as $R_{2}$. Similarly, we have
    \begin{equation*}
        d_{Q} \left ( R_{0},\gamma_{(R_{0},R_{0}\hat{\boldsymbol{w}}_{0})}(1) \right ) = \sqrt{\boldsymbol{w}_{0}^{\top}Q\boldsymbol{w}_{0}} < r.
    \end{equation*}
    This implies $\boldsymbol{w}_{0} \in B_{Q}(\boldsymbol{0},r)$, allowing us to state that $\mathcal{R}_{G}(\mathcal{I}_{G},1)\subset \pi(\mathcal{B}_{Q}(R_{0},r))$.
    
    Therefore, we have $\mathcal{B}_{Q}(R_{0},r) = \pi(\mathcal{R}_{G}(\mathcal{I}_{G},1))$. \hfill ~\qed
\end{pf}

Note that the initial set $\mathcal{I}_{G}$ is the Cartesian product of a fixed point $R_{0}$ in $\mathrm{SO(3)}$ and a metric ball $B_{Q}(\boldsymbol{0},r)$ in $\mathbb{R}^{3}$. This enables us to represent it as a compact and convex set in Euclidean charts, enabling the use of Euclidean reachability methods to characterize the mapped region of $\mathcal{B}_{Q}$ within these charts.

\subsection{Representation in the exp altas}

To illustrate the effectiveness of the equivalent representation of metric balls on $(\mathrm{SO}(3),\mathbb{G}_{Q})$ given in Proposition~\ref{Prop:SO3_Ball}, we present a specific attitude representation in which the metric balls can be represented as reachable sets in Euclidean space.

According to~\cite[Definition 8.1]{gallier2020differential}, a chart $\Psi$ is a diffeomorphism between $\mathcal{X} \subset \mathrm{SO(3)}$ and its image $\Psi(\mathcal{X}) \subset \mathbb{R}^{3}$. An atlas is a collection of pairs $\left\{(\mathcal{X}_{i},\Psi_{i})\right\}$ such that the union of the chart domain covers the entire $\mathrm{SO(3)}$ and every transition map between two overlapping regions of charts is smooth. An atlas with the minimum number of charts is known as a minimal atlas. Here, we present a practical minimal atlas to represent these metric balls in $\mathrm{SO(3)}$. Inspired by the work from~\cite{grafarend2011minimal}, this minimal atlas is called \textit{exp atlas} and is defined by 
\begin{align*}
    &\mathcal{X}_{i} = \left \{ R \in \mathrm{SO}(3) \mid \rm{tr}(\Lambda(\boldsymbol{s}_{i})R) \neq -1 \right \}, \\
    &\Psi_{i}(R) = (\log(\Lambda(\boldsymbol{s}_{i})R))^{\vee}, \qquad i \in \left\{ 0,1,2,3 \right\},
\end{align*}
where $\tr(\cdot)$ is the trace function and the vectors $\boldsymbol{s}_{i}\in \mathbb{R}^3$ are defined as:
\begin{align*}
    \begin{aligned}
        \boldsymbol{s}_{0} &= [1,1,1]^{\top},\qquad &\boldsymbol{s}_{1} = [1,-1,-1]^{\top},\\
        \boldsymbol{s}_{2} &= [-1,1,-1]^{\top},\qquad &\boldsymbol{s}_{3} = [-1,-1,1]^{\top}.
    \end{aligned}
\end{align*}
Note that the inverse map for exp atlas is given by 
\begin{align*}
    &\Psi_{i}^{-1}(\boldsymbol{r}_{i}) = \Lambda(\boldsymbol{s}_{i})e^{\hat{\boldsymbol{r}}_{i}}, \qquad i \in \left\{ 0,1,2,3 \right\},\\
    &\Psi_{i}(\mathcal{X}_{i}) = \left \{ \boldsymbol{r}_{i} \in \mathbb{R}^{3} \mid \left \| \boldsymbol{r}_{i} \right \|  < \pi \right \}.
\end{align*}

\begin{lem}[Geodesic dynamics in exp atlas]\label{lem:geodesic_dynamic_incharts}
    Let the curve $\boldsymbol{r}_{i}: U \to \mathbb{R}^{3}$ represents the mapped geodesic in any chart $(\mathcal{X}_{i},\Psi_{i})$. Its dynamics can then be expressed in $\mathbb{R}^{6}$ as:
    \begin{align}\label{eq:geo_system_polar_coord}
        \begin{split}
            \dot{\boldsymbol{r}}_{i} &= f(\boldsymbol{r}_{i})\boldsymbol{w},\\
            \dot{\boldsymbol{w}} &= -Q^{-1}\hat{\boldsymbol{w}}Q\boldsymbol{w},
        \end{split}
    \end{align}
    where $f(\boldsymbol{r}_{i})$ satisfies
    \begin{align*}
        &f(\boldsymbol{r}_{i}) = I + \frac{1}{2}\hat{\boldsymbol{r}}_{i} +  \frac{\hat{\boldsymbol{r}}_{i}^{2}}{\left \| \boldsymbol{r}_{i} \right \|^{2}} - \frac{\sin(\left \| \boldsymbol{r}_{i} \right \|)}{2 \left \| \boldsymbol{r}_{i} \right \|(1-\cos(\left \| \boldsymbol{r}_{i} \right \|))}\hat{\boldsymbol{r}}_{i}^{2}.
    \end{align*}
\end{lem}
\begin{pf}
    For each reduced geodesic $\boldsymbol{r}_{i}(t)$, its corresponding geodesic on $\mathrm{SO}(3)$ satisfies $R(t) = \Lambda(\boldsymbol{s}_{i})e^{\hat{\boldsymbol{r}}_{i}(t)}$. By applying the derivative property of the exponential map~\cite[Theorem 5.4]{hall2013lie}, we have 
    \begin{align*}
        \frac{\mathrm{d}}{\mathrm{d} t}R(t) &= \frac{\mathrm{d}}{\mathrm{d} t} \Lambda(\boldsymbol{s}_{i})e^{\hat{\boldsymbol{r}}_{i}(t)} = \Lambda(\boldsymbol{s}_{i})\frac{\mathrm{d}}{\mathrm{d} t} e^{\hat{\boldsymbol{r}}_{i}(t)}\\
        & = \Lambda(\boldsymbol{s}_{i})e^{\hat{\boldsymbol{r}}_{i}(t)}\left \{ \sum_{k=0}^{\infty} (-1)^{k}\frac{\mathrm{ad}_{\hat{\boldsymbol{r}}_{i}}^{k}\left ( \hat{\dot{\boldsymbol{r}}}_{i} \right )}{(k+1)!}\right \}.
    \end{align*}
    where $\mathrm{ad}$ is the adjoint operator of $\mathfrak{so}(3)$ defined by $\mathrm{ad}_{\hat{\boldsymbol{x}}}(\hat{\boldsymbol{y}}) = [\hat{\boldsymbol{x}},\hat{\boldsymbol{y}}]$, for any $\boldsymbol{x},\boldsymbol{y} \in \mathbb{R}^{3}$. From the geodesic equations~\eqref{eq:geo_dynamic} we have
    \begin{align*}
        \boldsymbol{w} =&(R^{\top}\dot{R})^{\vee} = \left \{ \sum_{k=0}^{\infty} (-1)^{k}\frac{\mathrm{ad}_{\hat{\boldsymbol{r}}_{i}}^{k}\left ( \hat{\dot{\boldsymbol{r}}}_{i} \right )}{(k+1)!}\right \}^{\vee}\\
        = & \left \{ \hat{\dot{\boldsymbol{r}}}_{i} - \frac{[\hat{\boldsymbol{r}}_{i},\hat{\dot{\boldsymbol{r}}}_{i}]}{2!} + \frac{[\hat{\boldsymbol{r}}_{i},[\hat{\boldsymbol{r}}_{i},\hat{\dot{\boldsymbol{r}}}_{i}]]}{3!} + \ldots \right \}^{\vee}\\
        = &  \dot{\boldsymbol{r}}_{i} - \frac{\hat{\boldsymbol{r}}_{i}\dot{\boldsymbol{r}}_{i}}{2!} + \frac{\hat{\boldsymbol{r}}_{i}\hat{\boldsymbol{r}}_{i}\dot{\boldsymbol{r}}_{i}}{3!} + \ldots \\
        = & \left (I - \sum_{k=1}^{+\infty } \frac{\hat{\boldsymbol{r}}^{2k-1}_{i}}{(2k)!} + \sum_{k=1}^{+\infty } \frac{\hat{\boldsymbol{r}}^{2k}_{i}}{(2k+1)!}  \right )\dot{\boldsymbol{r}}_{i} \\
        = & \left (I + \frac{\cos(\left \| \boldsymbol{r}_{i} \right \|) -1 }{\left \| \boldsymbol{r}_{i} \right \| ^2}\hat{\boldsymbol{r}}_{i} + \frac{ \left \| \boldsymbol{r}_{i} \right \| - \sin(\left \| \boldsymbol{r}_{i} \right \|)}{\left \| \boldsymbol{r}_{i} \right \| ^3}\hat{\boldsymbol{r}}^{2}_{i} \right )\dot{\boldsymbol{r}}_{i}.
    \end{align*}
    where the third equality holds because $ \left(\mathrm{ad}_{\hat{\boldsymbol{x}}}(\hat{\boldsymbol{y}})\right)^{\vee}= \hat{\boldsymbol{x}}\boldsymbol{y}$, for any $\boldsymbol{x},\boldsymbol{y} \in \mathbb{R}^{3}$. 
    Let $\dot{\boldsymbol{r}}_{i} = f(\boldsymbol{r}_{i})\boldsymbol{w}$ and the fifth equality holds because $\hat{\boldsymbol{x}}^{(2k-1)} = (-1)^{(k-1)}\|\boldsymbol{x}\|^{(2k-2)}\hat{\boldsymbol{x}}$, for every $\boldsymbol{x}\in \mathbb{R}^3$. It can be verified that, for any $\boldsymbol{x} \in \mathbb{R}^{3}$, the following identity holds: 
    \begin{align*}
        f(\boldsymbol{x}) =&  I + \frac{1}{2}\hat{\boldsymbol{x}} +  \frac{\hat{\boldsymbol{x}}^{2}}{\left \| \boldsymbol{x} \right \|^{2}} - \frac{\sin(\left \| \boldsymbol{x} \right \|)}{2 \left \| \boldsymbol{x} \right \|(1-\cos(\left \| \boldsymbol{x} \right \|))}\hat{\boldsymbol{x}}^{2} \\ = & \left (I + \frac{\cos(\left \| \boldsymbol{x}\right \|) -1 }{\left \| \boldsymbol{x} \right \| ^2}\hat{\boldsymbol{x}} + \frac{ \left \| \boldsymbol{x} \right \| - \sin(\left \| \boldsymbol{x} \right \|)}{\left \| \boldsymbol{x} \right \| ^3}\hat{\boldsymbol{x}}^{2} \right )^{-1}.
    \end{align*}
    Therefore, the dynamic equations can be represented as
    \begin{align*}
        \dot{\boldsymbol{r}}_{i} &= f(\boldsymbol{r}_{i})\boldsymbol{w},\\
        \dot{\boldsymbol{w}} &= -Q^{-1}\hat{\boldsymbol{w}}Q\boldsymbol{w}.
    \end{align*}
    This completes the proof.\hfill ~\qed
\end{pf}

Having proven that the geodesic dynamics in any exp chart $(\mathcal{X}_{i},\Psi_{i})$ satisfy Equations~\eqref{eq:geo_system_polar_coord}, we can represent all arguments and regions on $\mathrm{SO}(3)$ in the Proposition~\ref{Prop:SO3_Ball} within the chart. Based on this, we present the following theorem, which is essentially the mapped version of Proposition~\ref{Prop:SO3_Ball} in this chart.

\begin{thm}[Euclidean representation]
    Suppose $\mathcal{B}_{Q}(R_{0},r) \subset \mathcal{X}_{i}$. Let $\Psi_{i}(R_{0}) = \boldsymbol{r}_{i}^{0}$. Define $\mathcal{R}_{E}(\Theta,T)$ as the reachable set of the system described by~\eqref{eq:geo_system_polar_coord} at time $T$, initiating from $\Theta = \boldsymbol{r}_{i}^{0} \times B_{Q}(\boldsymbol{0},r)$. It holds that 
    \begin{align}
        \Psi_{i}(\mathcal{B}_{Q}(R_{0},r)) = \pi_{\boldsymbol{r}}(\mathcal{R}_{E}(\Theta,1)),
    \end{align}
    where $\pi_{\boldsymbol{r}}:\mathbb{R}^{6} \to \mathbb{R}^{3}$ is the projection defined by $\pi_{\boldsymbol{r}}([\boldsymbol{r};\boldsymbol{w}]) = \boldsymbol{r}$.
\end{thm}
\begin{rem}[Representation of metric balls]\;
\begin{enumerate}
    \item The projection of reachable set $\mathcal{R}_{E}$ can be considered as an Euclidean representation of the metric ball $\mathcal{B}_{Q}(R_{0},r)$ in $\mathbb{R}^{3}$. This representation allows computations on this metric ball using existing Euclidean reachability techniques. 
    \item In the context above, we assume that the metric ball is fully contained within a single exp chart. However, there are situations where the metric ball may not fit within a single chart and must be represented across multiple charts. We leave this issue for future work.
\end{enumerate}   
\end{rem}

\section{Numerical Results}\label{Sec:Num_examples}
\begin{figure*}[t]
    \centering
    \begin{subfigure}[b]{0.45\textwidth} 
        \includegraphics[width=\textwidth]{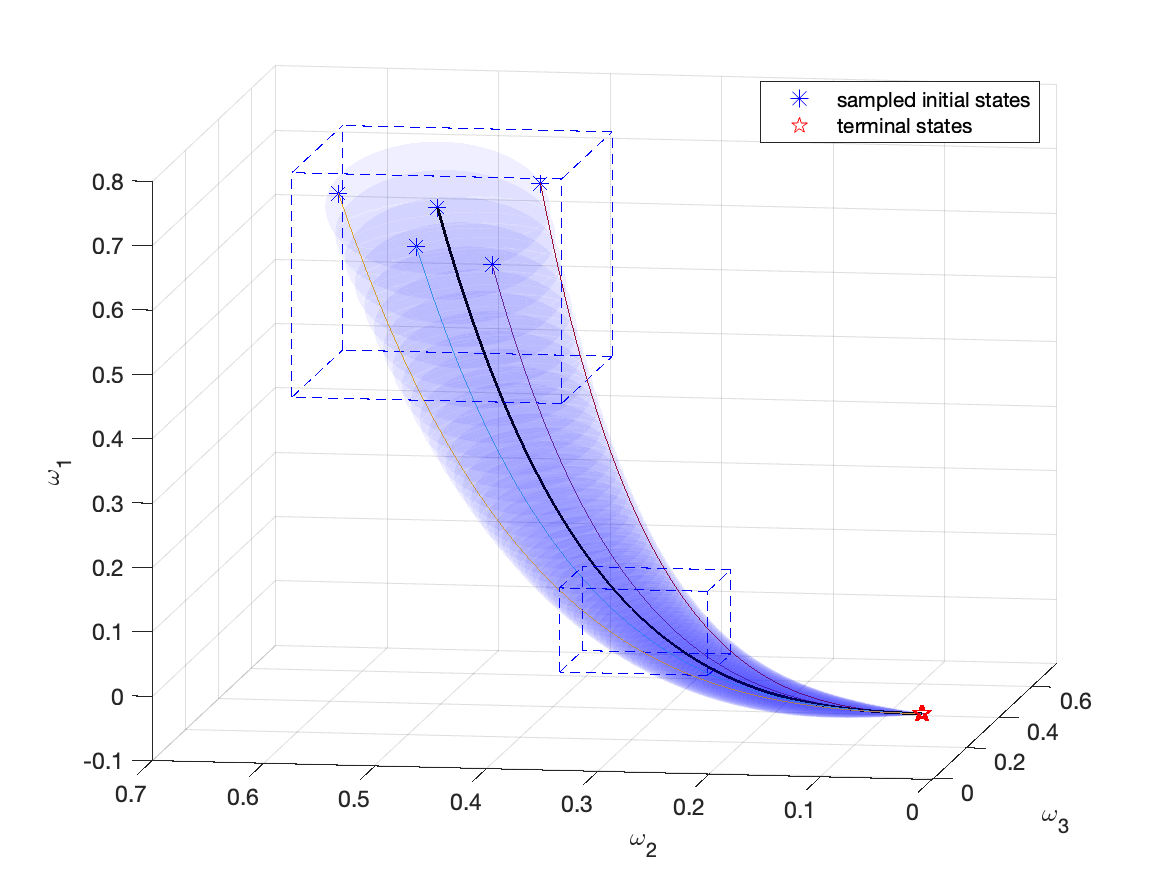}
        \caption{}
        \label{subfig1:ReachR3}
    \end{subfigure}
    \hfill
    \begin{subfigure}[b]{0.45\textwidth}
        \includegraphics[width=\textwidth]{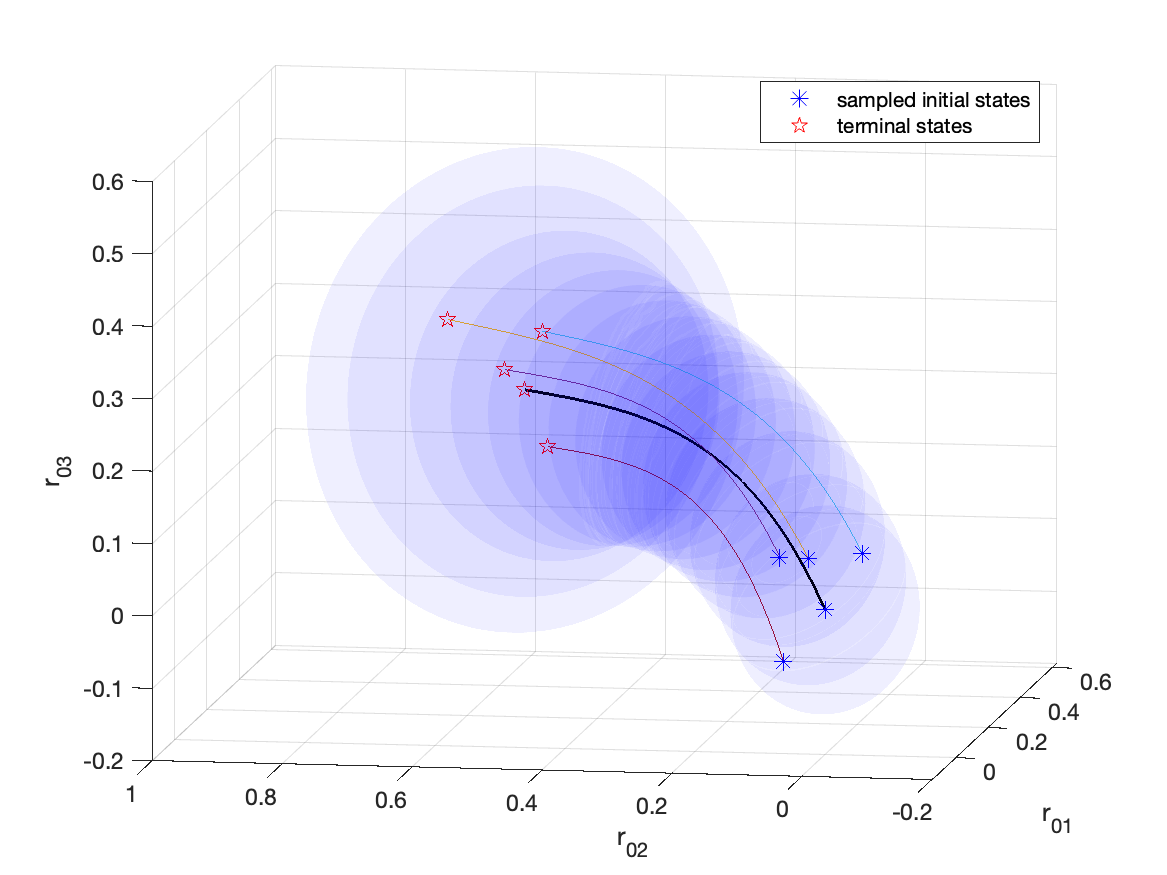}
        \caption{}
        \label{subfig2:ReachSO3}
    \end{subfigure}
    \caption{Over-approximated reachable sets for attitude control systems with the initial set $\mathcal{I}$ over the time interval $[0,4]$.
    (a) Fast over-approximation of the reachable set for angular velocity $\boldsymbol{\omega} \in \mathbb{R}^{3}$ (blue region) using the Euclidean reachability algorithm LDF2. The dashed boxes represent over-approximated reachable tubes during $[t_{0}=0, t_{1}=0.1]$ and $[t_{6}=0.6, t_{7}=0.7]$, which serve as search regions in our ConReach algorithm at Step 0 and Step 6, respectively.  
    (b) Over-approximation of reachable set of attitudes (blue region) computed using our ConReach algorithm and represented in the exp chart $(\mathcal{X}_{0}, \Psi_{0})$.}
    \label{fig:results}
\end{figure*}

In this section, we illustrate our reachability methods for an attitude control system~\eqref{eq:attitude_dynamic2} with a specific choice of feedback controller $\boldsymbol{\tau}=\boldsymbol{\tau}(R,\boldsymbol{\omega})$. 
Consider the attitude control system~\eqref{eq:attitude_dynamic2} with the inertia matrix given by $J = \mathrm{diag}(-2,-1,-3)$ and control input $\boldsymbol{\tau}$ defined by 
\begin{equation*}
    \boldsymbol{\tau}= J^{2}\boldsymbol{\omega} +\hat{\boldsymbol{\omega}} J\boldsymbol{\omega}.
\end{equation*}
This control law is designed to stabilize the system's angular velocity, ensuring that every trajectory converges to the equilibrium submanifold $\mathcal{S} = \mathrm{SO}(3)\times \{\boldsymbol{0}_3\}$.
Consider an initial set $\mathcal{I} = \mathcal{B}_{I}(R_{0}, r_{R}) \times B_{I}(\boldsymbol{\omega}_{0},r_{V})$ with 
\begin{equation*}
    \begin{aligned}
        &R_{0} = I,  &r_{R} = 0.1, \\
        &\boldsymbol{\omega}_{0} = [0.65,0.54,0.61]^{\top}, \qquad & r_{V} = 0.1.
    \end{aligned}
\end{equation*}
We use the proposed ConReach algorithm~\eqref{alg:main} to obtain an over-approximation of the reachable set at the terminal time $T = 4$. 

In this example, the mapping $X_{\boldsymbol{\omega}}(R,\boldsymbol{\omega})$ depends solely on the angular velocity $\boldsymbol{\omega}$, and is given by $X_{\boldsymbol{\omega}} = J\boldsymbol{\omega}$.
As a result, at each step $i$, the search region $\mathcal{U}_{i}$ can be simplified to the over-approximated reachable tube of the angular velocity system in $\mathbb{R}^{3}$ over the time interval $[t_{i},t_{i+1}]$. To compute $\mathcal{U}_{i}$, we apply the LDF2 algorithm proposed in~\cite{fan2017simulation}, which provides a fast over-approximation of the Euclidean reachable sets. The search region $\mathcal{U}_{i}$ is then computed as a bounding box that contains all the reachable sets of the angular velocity over each time interval $[t_{i},t_{i+1}]$.

The results of this computation are illustrated in Figure~\ref{subfig1:ReachR3}. 
Several colored lines represent different simulations randomly initiated from the initial set $\mathcal{I}$. These simulations help demonstrate the validity of the over-approximated reachable set for $\boldsymbol{\omega}$ and the bounding box $\mathcal{U}_{i}$. 
All angular velocity trajectories converge to zero (the equilibrium point), validating the design of the control inputs.

\begin{figure*}[ht]
    \centering
    \begin{equation}\label{prob:optimization_example}
        \begin{aligned}
            \min \quad &  -\tr(Q_{i+1})\\
            \textrm{s.t.} 
            \quad & Q_{i+1} \succeq 0, \quad Q_{i+1}\preceq Q_{i}, \quad P_{i+1} \succeq 0, \quad P_{i+1}\preceq P_{i},\\
            \quad & \begin{bmatrix}
                \hat{\boldsymbol{\omega}}Q_{i+1} - Q_{i+1}\hat{\boldsymbol{\omega}} - 2c_{i+1}Q_{i+1}  & Q_{i+1}\\
                Q_{i+1} & J^{\top}P_{i+1} + P_{i+1}J - 2c_{i+1}P_{i+1} 
              \end{bmatrix} \preceq 0, \quad \forall \hat{\boldsymbol{\omega}} \in \mathcal{O}_{i}, \\
        \end{aligned}
    \end{equation}

    \begin{align}\label{sol:SDP}
    \begin{split}
    &\mathrm{Step \  0:} \qquad c_{1} = 0.1871,  \qquad Q_{1} =\begin{bmatrix}
        0.989& -0.020 & 0.010\\
        -0.020& 0.962& 0.019 \\
        0.010&  0.019& 0.991
      \end{bmatrix}, \qquad P_{1} =\begin{bmatrix}
        0.710& 0.017 & -0.015\\
        0.017& 0.999& 0.015\\
        -0.015&  0.015& 0.491
      \end{bmatrix}, \\
    &\mathrm{Step \ 6:} \qquad c_{7} = 0.1871, \qquad Q_{7} =\begin{bmatrix}
        0.987& -0.025 & 0.013\\
        -0.025& 0.909& 0.039 \\
        0.013&  0.039& 0.983
    \end{bmatrix}, \qquad P_{7} =\begin{bmatrix}
        0.670& 0.010 & 0.012\\
        0.010& 0.997& 0.019\\
        0.012&  0.019& 0.464
      \end{bmatrix}.
    \end{split}
    \end{align}
\end{figure*}
We present the computation results for the search region $\mathcal{U}_{i}$ at step 0 and step 6:
\begin{align*}   
    & \mathcal{U}_{0} = [0.442,0.750] \times [0.398,0.640] \times [0.361,0.710],\\
    &\mathcal{U}_{6} = [0.110,0.250] \times [0.218,0.351] \times [0.025,0.155].
\end{align*}
We depict these two bounding boxes using dotted lines in Figure~\ref{subfig1:ReachR3}.

In this example, the interval matrix $\mathcal{A}_{i}$ is simplified to $\boldsymbol{0}_{3 \times 3}$, while $\mathcal{B}_{i}$ is set to the constant matrix $J$. The interval matrix $\mathcal{O}_{i}$ is chosen such that $\hat{\boldsymbol{\omega}} \in \mathcal{O}_{i}$ for all $\boldsymbol{\omega} \in \mathcal{U}_{i}$. With the value of $c_{i+1}$ fixed during the line search, the SDP problem for each step is formulated as Problem~\eqref{prob:optimization_example}.
We present the solution to the SDP problem at Step 0 and Step 6 with $c_{1} = c_{7} = 0.1871$ in Equation~\eqref{sol:SDP}. 
The resulting over-approximated reachable sets of attitude in the exp chart $(\mathcal{X}_{0},\Psi_{0})$ are plotted at multiple time steps in Figure~\ref{subfig2:ReachSO3}. 
All colored lines represent rotation trajectories, following the color scheme defined in Figure~\ref{subfig1:ReachR3}.
We implement the ConReach algorithm in MATLAB. Solving these two reachability problems with $N = 40$ and $N_s = 3$ takes 14.73s on a MacBook Air with an Apple M2 chip and 8 GB of RAM. YALMIP~\cite{toh1999sdpt3} and SDPT3~\cite{lofberg2004yalmip} are used as solvers for SDP problems.

\section{Conclusion and Future works}\label{Sec:Conclusion}

In this paper, we leverage contraction theory on Riemannian manifolds to develop a novel simulation-based approach for the reachability analysis of attitude control systems. Central to our approach is a parametrized family of Riemannian metrics on the manifold $\mathrm{SO}(3) \times \mathbb{R}^{3}$. By characterizing the incremental distance between trajectories of the attitude system with respect to this family of metrics, our method enables an efficient search for the optimal metric via semidefinite programming (SDP).
Using these incremental trajectory distances, we establish theoretical guarantees for over-approximations of the reachable sets of attitude dynamics. Furthermore, we provide a practical representation of these over-approximations, ensuring they can be visualized and analyzed using standard Euclidean tools and software.
For future work, we plan to explore applications in safe control synthesis and path planning. Additionally, we aim to extend our reachability analysis to systems that couple both attitude and position dynamics.

\bibliographystyle{ieeetr}        
\bibliography{main}           

\appendix 
\section{Properties of the Coefficient Vector $\boldsymbol{\varsigma}_{i,j}$}\label{Sec:App_Coeff_Vec}
Let $(\mathrm{SO(3)}, \mathbb{G}_{Q})$ be a Riemannian manifold, and let $\{X_{L_{1}},X_{L_{2}},X_{L_{3}}\}$ denote the basis for $\mathcal{X}(\mathrm{SO(3)})$. For any $i,j \in \left \{ 1,2,3 \right \}$, the covariant derivative of $X_{L_{i}}$ and $X_{L_{j}}$ is represented by ${\overset{R}{\nabla}}_{X_{L_{i}}}X_{L_{j}}(R) =  \sum_{k = 1}^{3} \varsigma^{k}_{i,j}X_{L_{k}}(R) = R \hat{\boldsymbol{\varsigma}}_{i,j}$.
The coefficient $\boldsymbol{\varsigma}_{i,j}$ has the following properties:
\begin{enumerate}
    \item\label{varsgm_prop_1}$\boldsymbol{\varsigma}_{i,j} = Q^{-1}\boldsymbol{\eta}_{i,j}$, where
    \begin{align*}
        \boldsymbol{\eta}_{i,j}  =& \left [ \eta^{1}_{i,j}, \eta^{2}_{i,j}, \eta^{3}_{i,j}\right ]^{\top}, \\
        \eta^{l}_{i,j} =& -\frac{1}{2}\langle L_{i},[L_{j},L_{l}] \rangle_{\mathbb{G}_{Q}(I)} -\frac{1}{2} \langle L_{j},[L_{i},L_{l}] \rangle_{\mathbb{G}_{Q}(I)} \notag \\
        &-\frac{1}{2} \langle L_{l},[L_{j},L_{i}] \rangle_{\mathbb{G}_{Q}(I)}, \qquad \forall l \in \left \{ 1,2,3 \right \}. \notag
    \end{align*}
    \item\label{varsgm_prop_2} for any $\boldsymbol{\omega},\boldsymbol{\alpha} \in \mathbb{R}^{3}$,
    \begin{equation*}
    \sum_{i,j = 1}^{3}\alpha_{i}\omega_{j}\boldsymbol{\varsigma}_{i,j} = Q^{-1} \hat{\boldsymbol{\omega}}Q\boldsymbol{\alpha}-\frac{1}{2}\tr(Q) Q^{-1}\hat{\boldsymbol{\omega}}\boldsymbol{\alpha}.
    \end{equation*}
\end{enumerate}
    
\begin{pf}\label{pf:covar_deriv_coeff}
~\eqref{varsgm_prop_1} We start by introducing an inner product function $\eta^{l}_{i,j}: \mathrm{SO}(3) \to \mathbb{R}$ for each $i,j,l \in \left \{ 1,2,3 \right \}$, defined by $\eta^{l}_{i,j} = \langle {\nabla}_{X_{L_{i}}}X_{L_{j}},X_{L_{l}} \rangle_{\mathbb{G}_{Q}}$.
Using the properties of the Levi-Civita connection in~\eqref{Levi_Connection_prop}, we obtain
    \begin{align}\label{eq:eta_lij}
        \eta^{l}_{i,j} = & \langle {\nabla}_{X_{L_{i}}}X_{L_{j}},X_{L_{l}} \rangle_{\mathbb{G}_{Q}}  \notag\\
        = & \frac{1}{2}   \left ( -  \langle X_{L_{i}},[X_{L_{j}},X_{L_{l}}] \rangle_{\mathbb{G}_{Q}} - \langle X_{L_{j}},[X_{L_{i}},X_{L_{l}}] \rangle_{\mathbb{G}_{Q}} \right . \notag\\
            & \quad \left . - \langle X_{L_{l}},[X_{L_{j}},X_{L_{i}}] \rangle_{\mathbb{G}_{Q}} + X_{L_{i}}\langle X_{L_{j}},X_{L_{l}} \rangle_{\mathbb{G}_{Q}}  \right . \notag\\
            &\quad \left .  + X_{L_{j}}\langle X_{L_{i}},X_{L_{l}} \rangle_{\mathbb{G}_{Q}}- X_{L_{l}}\langle X_{L_{i}},X_{L_{j}} \rangle_{\mathbb{G}_{Q}} \right ) \notag\\
        = & \frac{1}{2} \left ( - \langle X_{L_{i}},[X_{L_{j}},X_{L_{l}}] \rangle_{\mathbb{G}_{Q}} - \langle X_{L_{j}},[X_{L_{i}},X_{L_{l}}] \rangle_{\mathbb{G}_{Q}} \right . \notag\\
            & \quad \left . - \langle X_{L_{l}},[X_{L_{j}},X_{L_{i}}] \rangle_{\mathbb{G}_{Q}} \right ) \notag \\
        =& -\frac{1}{2}\langle L_{i},[L_{j},L_{l}] \rangle_{\mathbb{G}_{Q}(I)} -\frac{1}{2} \langle L_{j},[L_{i},L_{l}] \rangle_{\mathbb{G}_{Q}(I)} \notag\\
            &-\frac{1}{2} \langle L_{l},[L_{j},L_{i}] \rangle_{\mathbb{G}_{Q}(I)}.
    \end{align}
    The second equality above follows from the observation that for each $i,j,l \in \left \{ 1,2,3 \right \}$, the inner product $\langle X_{L_{i}},X_{L_{j}} \rangle_{\mathbb{G}_{Q}}: \mathrm{SO}(3) \to \mathbb{R} $ remains constant, leading to $X_{L_{i}}\langle X_{L_{j}},X_{L_{l}}\rangle_{\mathbb{G}_{Q}} (R)  =  0$. The third equality follows from the property that the bracket of two left-invariant vector fields is also left-invariant~\cite[Chap. 1]{do1992riemannian}. Thus, $\eta^{l}_{i,j}$ also remains constant on $\mathrm{SO}(3)$. 
    
    Define $\boldsymbol{\eta}_{i,j} = \left [ \eta^{1}_{i,j}, \eta^{2}_{i,j}, \eta^{3}_{i,j}\right ]^{\top}$. It can be inferred that 
    \begin{align}
        \boldsymbol{\eta}_{i,j}  &= \begin{bmatrix}
            \langle {\nabla}_{X_{L_{i}}}X_{L_{j}},X_{L_{1}} \rangle_{\mathbb{G}_{Q}}  \\
            \langle {\nabla}_{X_{L_{i}}}X_{L_{j}},X_{L_{2}} \rangle_{\mathbb{G}_{Q}} \\
            \langle {\nabla}_{X_{L_{i}}}X_{L_{j}},X_{L_{3}} \rangle_{\mathbb{G}_{Q}}
           \end{bmatrix} \notag \\
           & = \begin{bmatrix}
            \sum_{k = 1}^{3}\varsigma^{k}_{i,j}\langle X_{L_{k}},X_{L_{1}} \rangle_{\mathbb{G}_{Q}}  \\
            \sum_{k = 1}^{3}\varsigma^{k}_{i,j}\langle X_{L_{k}},X_{L_{2}}  \rangle_{\mathbb{G}_{Q}} \\
            \sum_{k = 1}^{3}\varsigma^{k}_{i,j}\langle X_{L_{k}},X_{L_{3}} \rangle_{\mathbb{G}_{Q}}
           \end{bmatrix} \notag \\
          &  = \begin{bmatrix}
            \langle L_{1},L_{1} \rangle_{\mathbb{G}_{Q}(I)} &\langle L_{2},L_{1} \rangle_{\mathbb{G}_{Q}(I)} &\langle L_{3},L_{1} \rangle_{\mathbb{G}_{Q}(I)}  \\
            \langle L_{1},L_{2} \rangle_{\mathbb{G}_{Q}(I)} &\langle L_{2},L_{2} \rangle_{\mathbb{G}_{Q}(I)} &\langle L_{3},L_{2} \rangle_{\mathbb{G}_{Q}(I)} \\
            \langle L_{1},L_{3} \rangle_{\mathbb{G}_{Q}(I)} &\langle L_{2},L_{3} \rangle_{\mathbb{G}_{Q}(I)} &\langle L_{3},L_{3} \rangle_{\mathbb{G}_{Q}(I)}
           \end{bmatrix} \boldsymbol{\varsigma}_{i,j} \notag \\
           &= Q\boldsymbol{\varsigma}_{i,j}.
    \end{align}
Since $Q$ is positive definite, it is also invertible, and thus we obtain $\boldsymbol{\varsigma}_{i,j} = Q^{-1}\boldsymbol{\eta}_{i,j}$.

~\eqref{varsgm_prop_2}  As shown in Equation~\eqref{eq:eta_lij}, for a fixed $Q$, the coefficient vector $\boldsymbol{\eta}_{i,j}$ for each $i,j \in \left \{ 1,2,3 \right \}$ can be computed using Equation~\eqref{eq:eta_lij}, resulting in
\begin{equation} \label{eq:Q_value}
\begin{aligned}
    \boldsymbol{\eta}_{1,1} &= \left [ 0, -Q_{13},Q_{12}\right ]^{\top}, \\
    \boldsymbol{\eta}_{1,2} &= \left [ Q_{13}, 0, (-Q_{11}+Q_{22}+Q_{33})/2\right ]^{\top}, \\
    \boldsymbol{\eta}_{1,3} &= \left [ -Q_{12}, (Q_{11}-Q_{22}-Q_{33})/2, 0\right ]^{\top}, \\
    \boldsymbol{\eta}_{2,1} &= \left [ 0,-Q_{23},(-Q_{11}+Q_{22}-Q_{33})/2\right ]^{\top}, \\
    \boldsymbol{\eta}_{2,2} &= \left [ Q_{23},0,-Q_{12}\right ]^{\top}, \\
    \boldsymbol{\eta}_{2,3} &= \left [ (Q_{11}-Q_{22}+Q_{33})/2,Q_{12},0\right ]^{\top}, \\
    \boldsymbol{\eta}_{3,1} &= \left [ 0,(Q_{11}+Q_{22}-Q_{33})/2,Q_{23}\right ]^{\top}, \\
    \boldsymbol{\eta}_{3,2} &= \left [ (-Q_{11}-Q_{22}+Q_{33})/2,0,-Q_{13}\right ]^{\top}, \\
    \boldsymbol{\eta}_{3,3} &= \left [ -Q_{23},Q_{13},0\right ]^{\top}.
\end{aligned}
\end{equation}
Then, it holds that
    \begin{align*}
        &\sum_{i,j = 1}^{3}\alpha_{i}\omega_{j}\boldsymbol{\varsigma}_{i,j} = \sum_{i,j = 1}^{3}\alpha_{i}\omega_{j}Q^{-1}\boldsymbol{\eta}_{i,j} = Q^{-1}\sum_{i,j = 1}^{3}\alpha_{i}\omega_{j}\boldsymbol{\eta}_{i,j} \\
        & =Q^{-1} \begin{bmatrix}
            \sum_{j = 1}^{3} \omega_{j}\eta^{1}_{1,j} & \sum_{j = 1}^{3} \omega_{j}\eta^{1}_{2,j} & \sum_{j = 1}^{3} \omega_{j}\eta^{1}_{3,j}\\
            \sum_{j = 1}^{3} \omega_{j}\eta^{2}_{1,j} & \sum_{j = 1}^{3} \omega_{j}\eta^{2}_{2,j} & \sum_{j = 1}^{3} \omega_{j}\eta^{2}_{3,j} \\
            \sum_{j = 1}^{3} \omega_{j}\eta^{3}_{1,j} & \sum_{j = 1}^{3} \omega_{j}\eta^{3}_{2,j} & \sum_{j = 1}^{3} \omega_{j}\eta^{3}_{3,j}
          \end{bmatrix} \boldsymbol{\alpha} \\
          &=Q^{-1}A\boldsymbol{\alpha}.
    \end{align*}
        Applying Equation~\eqref{eq:Q_value}, the matrix $A$ satisfies that
    \begin{align*}
        A = &\begin{bmatrix}
            \sum_{j = 1}^{3} \omega_{j}\eta^{1}_{1,j} & \sum_{j = 1}^{3} \omega_{j}\eta^{1}_{2,j} & \sum_{j = 1}^{3} \omega_{j}\eta^{1}_{3,j}\\
            \sum_{j = 1}^{3} \omega_{j}\eta^{2}_{1,j} & \sum_{j = 1}^{3} \omega_{j}\eta^{2}_{2,j} & \sum_{j = 1}^{3} \omega_{j}\eta^{2}_{3,j} \\
            \sum_{j = 1}^{3} \omega_{j}\eta^{3}_{1,j} & \sum_{j = 1}^{3} \omega_{j}\eta^{3}_{2,j} & \sum_{j = 1}^{3} \omega_{j}\eta^{3}_{3,j}
          \end{bmatrix}\\
          =& \begin{bmatrix}
            \omega_{2}Q_{13}-\omega_{3}Q_{12}& \omega_{2}Q_{23}-\omega_{3}Q_{22} & \omega_{2}Q_{33}-\omega_{3}Q_{23}\\
            \omega_{3}Q_{11}-\omega_{1}Q_{13} & \omega_{3}Q_{12}-\omega_{1}Q_{23} & \omega_{3}Q_{13}-\omega_{1}Q_{33}\\
            \omega_{1}Q_{12}-\omega_{2}Q_{11} & \omega_{1}Q_{22}-\omega_{2}Q_{12} & \omega_{1}Q_{23}-\omega_{2}Q_{13}
          \end{bmatrix} \\
        & +\frac{1}{2}(Q_{11}+Q_{22}+Q_{33})\begin{bmatrix}
            0& \omega_{3} & -\omega_{2}\\
            -\omega_{3}& 0 & \omega_{1}\\
            \omega_{2}&  -\omega_{1}&0
          \end{bmatrix}\\
        = &\hat{\boldsymbol{\omega}}Q-\frac{1}{2}\tr(Q)\hat{\boldsymbol{\omega}}.
    \end{align*}
    We therefore obtain the equation
    \begin{align*}
        \sum_{i,j = 1}^{3}\alpha_{i}\omega_{j}\boldsymbol{\varsigma}_{i,j} & = Q^{-1}(\hat{\boldsymbol{\omega}}Q-\frac{1}{2}\tr(Q)\hat{\boldsymbol{\omega}})\boldsymbol{\alpha}\\
        &= Q^{-1} \hat{\boldsymbol{\omega}}Q\boldsymbol{\alpha}-\frac{1}{2}\tr(Q) Q^{-1}\hat{\boldsymbol{\omega}}\boldsymbol{\alpha}. 
    \end{align*}
    This completes the proof. \hfill ~\qed
\end{pf}

\section{Inclusion Relationship for Metric Balls}\label{Sec:App_Inclu_Relation}
We provide the proof of Lemma~\ref{lem:inclusion_product}.

\begin{pf}
    Let $\mathbb{R}^{3}$ be a Riemannian manifold. If $P_{1} \succeq P_{2}$, then for any $\boldsymbol{x}_{0},\boldsymbol{x} \in \mathbb{R}^{3}$, 
    \begin{align}\label{iqn:inclusion_R3}
       & d_{P_{2}}(\boldsymbol{x},\boldsymbol{x}_{0}) = \left \| \boldsymbol{x} - \boldsymbol{x}_{0} \right \|_{P_{2}} = \sqrt{(\boldsymbol{x} - \boldsymbol{x}_{0})^{\top}P_{2}(\boldsymbol{x} - \boldsymbol{x}_{0})} \notag \\
    \le &  \sqrt{(\boldsymbol{x} - \boldsymbol{x}_{0})^{\top}P_{1}(\boldsymbol{x} - \boldsymbol{x}_{0})} = d_{P_{1}}(\boldsymbol{x},\boldsymbol{x}_{0}).
    \end{align}
    The first inequality relies on the fact that for $P_{1} \succeq P_{2}$, it holds that $\boldsymbol{x}^{\top}P_{1}\boldsymbol{x} \ge \boldsymbol{x}^{\top}P_{2}\boldsymbol{x}$ for any $\boldsymbol{x} \in \mathbb{R}^{3}$.
    Let $\mathrm{SO(3)}$ be a Riemannian manifold. According to the Hopf-Rinow theorem, for any point $R_{1} \in \mathcal{B}_{Q_{1}}(R_{0},r)$, there exists a unique geodesic segment associated with $\mathbb{G}_{Q_{1}}$ that connects it to the center point $R_{0}$. Let this segment be denoted by $\gamma_{Q_{1}}:[0,1] \to \mathrm{SO}(3)$, with $\gamma_{Q_{1}}(0) = R_{0}$ and $\gamma_{Q_{1}}(1) = R_{1}$. If $Q_{1} \succeq Q_{2}$, then the length of $\gamma_{Q_{1}}$, with respect to $\mathbb{G}_{Q_{2}}$, satisfies 
    \begin{align*}
        \mathcal{L}_{\gamma_{Q_{1}}} &= \int_{0}^{1} \sqrt{\langle \dot{\gamma}_{Q_{1}}(t), \dot{\gamma}_{Q_{1}}(t)\rangle_{\mathbb{G}_{Q_{2}}(\gamma_{Q_{1}}(t))}} \d t \\
        & = \int_{0}^{1} \sqrt{\boldsymbol{w}^{\top}(t)Q_{2}\boldsymbol{w}(t)} \d t \\
        & \le \int_{0}^{1} \sqrt{\boldsymbol{w}^{\top}(t)Q_{1}\boldsymbol{w}(t)} \d t = d_{Q_{1}}(R_{0},R_{1}),
    \end{align*}
    where $\boldsymbol{w}(t) = (\gamma^{\top}_{Q_{1}}(t)\dot{\gamma}_{Q_{1}}(t))^{\vee}$ is the reduced curve of $\gamma_{Q_{1}}$.
    The distance between $R_{0}$ and $R_{1}$ with respect to $\mathbb{G}_{Q_{2}}$ represents the shortest path among all possible curves connecting these two points. Consequently, it must be less than or equal to the length of $\gamma_{Q_{1}}$, leading to the following relation:
    \begin{equation}\label{iqn:inclusion_SO3}
        d_{Q_{2}}(R_{0},R_{1}) \le \mathcal{L}_{\gamma_{Q_{1}}} \le d_{Q_{1}}(R_{0},R_{1}) < r.
    \end{equation}
Now let's consider the product manifold $\mathrm{SO(3)} \times \mathbb{R}^{3}$. Let $d_{i}$ denote the distance function on the metric space $(\mathrm{SO}(3) \times \mathbb{R}^3, \mathbb{G}_{Q_{i}} \times \mathbb{G}_{P_{i}})$, $i = \left\{ 1, 2 \right\}$. If $Q_{1} \succeq Q_{2}$ and $P_{1} \succeq P_{2}$, then for any $(R,\boldsymbol{\omega}) \in \mathrm{SO(3)} \times \mathbb{R}^{3}$, it holds that
\begin{align*}
    d_{2}((R,\boldsymbol{\omega}),(R_{0},\boldsymbol{\omega}_{0})) &= \sqrt{d^{2}_{Q_{2}}(R,R_{0}) + d^{2}_{P_{2}}(\boldsymbol{\omega},\boldsymbol{\omega}_{0})} \\
    & \le \sqrt{d^{2}_{Q_{1}}(R,R_{0}) + d^{2}_{P_{1}}(\boldsymbol{\omega},\boldsymbol{\omega}_{0})}\\
    & = d_{1}((R,\boldsymbol{\omega}),(R_{0},\boldsymbol{\omega}_{0})).
\end{align*}  
The first inequality follows from~\eqref{iqn:inclusion_R3} and~\eqref{iqn:inclusion_SO3}. Consequently, we can deduce that $\mathcal{B}_{Q_{1} \times P_{1}}((R_{0},\boldsymbol{\omega}_{0}),r) \subseteq \mathcal{B}_{Q_{2} \times P_{2}}((R_{0},\boldsymbol{\omega}_{0}),r)$.\hfill ~\qed 
\end{pf}

\end{document}